\documentclass[a4paper,reqno]{amsart}
\usepackage{amssymb}
\newtheorem{corollary}{Corollary}
\newtheorem{theorem}{Theorem}
\newtheorem{lemma}{Lemma}
\renewcommand{\thesection}{\arabic{section}}
\renewcommand{\theequation}{\thesection.\arabic{equation}}

\newcommand{\bfp}{{\bf p}}

\newcommand{\bfx}{{\bf x}}

\newcommand{\bfy}{{\bf y}}
\newcommand{\bfz}{{\bf z}}

\newcommand{\bfc}{{\bf c}}
\newcommand{\bfe}{{\bf e}}
\newcommand{\bfA}{{\bf A}}
\newcommand{\bfB}{{\bf B}}
\newcommand{\bfalpha}{\mbox{\boldmath $ \alpha$}}
\newcommand{\bfalphas}{\mbox{\scriptsize \boldmath $ \alpha$}}
\newcommand{\bfsigma}{\mbox{\boldmath $ \sigma$}}
\newcommand{\bfsigmas}{\mbox{\scriptsize \boldmath $ \sigma$}}
\newcommand{\bfxi}{\mbox{\boldmath $ \xi$}}

\newcommand{\bfomega}{\mbox{\boldmath $ \omega$}}
\newcommand{\bfomegas}{\mbox{\scriptsize \boldmath $ \omega$}}
\newcommand{\bfnabla}{\mbox{\boldmath $ \nabla$}}

\newcommand{\cA}{{\mathcal{A}}}
\newcommand{\cH}{{\mathcal{H}}}

\newcommand{\cO}{{\mathcal{O}}}

\newcommand{\ninfi}{{n \rightarrow \infty}}
\newcommand{\nin}{{n \in {\Bbb N}}}
\newcommand{\supp}{\mbox{supp}\,}

\begin{document}

\begin{center}
{\Large\bf Heat kernel estimates and spectral properties of a pseudorelativistic operator with magnetic field}
\end{center}

\vspace{0.5cm}

\begin{center}
{\large D.~H.~Jakubassa-Amundsen  }

Mathematics Institute,
University of Munich\\ Theresienstr. 39, 80333
Munich, Germany 
\end{center}

\vspace{1cm}

{\noindent\small
Based on the Mehler heat kernel of the  Schr\"{o}dinger operator  for a free electron in a constant
magnetic field an estimate for the kernel of $E_A= \,|\bfalpha (\bfp-e\bfA)+\beta m|\;$ is derived, where $E_A$ represents the kinetic energy of a Dirac electron within
the pseudorelativistic no-pair Brown-Ravenhall model.
This estimate is used to provide the bottom of the essential spectrum
for the two-particle Brown-Ravenhall operator, describing the motion of the electrons in a central Coulomb field and a constant magnetic field, if the central charge is restricted to $Z\leq 86.$
}
\vspace{0.2cm}

[J.~Math.~Phys. 49 (2008) 032305, 1-22]
\vspace{0.2cm}

{\noindent AMS: 81Q10}

\section{Introduction}

Consider two relativistic electrons of mass $m$ in an electromagnetic field which is
 generated by a point nucleus of charge $Z$ fixed at the origin, and by
a vector potential $\bfA.$

The two-particle Coulomb-Dirac operator, introduced by Sucher \cite{Su}
and accounting for a magnetic field is defined by
\begin{equation}\label{1.1a}
H_2\;=\;\sum_{k=1}^2 \left( D_A^{(k)}+V^{(k)}\right)\;+\;P_{+,2}\,V^{(12)}\,P_{+,2}
\end{equation}
$$D_A^{(k)}:=\;\bfalpha^{(k)}\;(\bfp_k-e\bfA(\bfx_k))\;+\;\beta^{(k)}m,\qquad k=1,2,$$
where $P_{+,2}$ projects onto the positive spectral subspace of $\sum_{k=1}^2 (D_A^{(k)}+V^{(k)})$.
The underlying Hilbert space is $\cA(L_2({\Bbb R}^3)\otimes {\Bbb C}^4)^2$
where $\cA$ denotes antisymmetrization with respect to electron exchange.
The single-particle and two-particle potentials are, respectively,
\begin{equation}\label{1.2a}
V^{(k)}\;=\;-\frac{\gamma}{x_k},\qquad V^{(12)}\;=\;\frac{e^2}{|\bfx_1-\bfx_2|}
\end{equation}
where the field strength $\gamma =Ze^2$ and $e^2 \approx 1/137.04$ the fine structure constant.
$x_k\,=\,|\bfx_k|$ is the modulus of the spatial coordinate of electron $k,\;\;k=1,2.$ The momentum of electron $k$ is denoted by $\bfp_k$, 
and $\beta,\;\;\bfalpha=(\alpha_1,\alpha_2,\alpha_3)$ and $\bfsigma=(\sigma_1,\sigma_2,\sigma_3)$ are the Dirac and Pauli matrices, respectively.

Due to the positron degrees of freedom the spectrum of the Coulomb-Dirac operator is unbounded from below.
In spectroscopic studies of static ions where pair creation plays no role one can instead
work with a semibounded operator, derived from $H_2$, which solely describes the electronic states.
One of the current techniques  to construct such an operator is by means of a (unitary)
Morse-Feshbach transformation scheme (see e.g. \cite{Kato,DeV,DK,Jaku05}) which
aims at decoupling the positive and negative spectral subspaces of the electron.
For $\bfA\neq {\bf 0}$ it is thereby crucial \cite{LSS} to include the vector potential in the definition of these 
 subspaces. A decoupling of the spectral subspaces up to second order in $e^2$  is provided in \cite{Jaku1}.
 
The first-order transformation leads to the  Brown-Ravenhall operator \cite{LSS}
\begin{equation}\label{1.3a}
H_2^{BR}\;=\;\sum_{k=1}^2 \Lambda_{A+,2}\,(D_A^{(k)}+V^{(k)})\,\Lambda_{A+,2}\;+\;\Lambda_{A+,2}\,V^{(12)}\,\Lambda_{A+,2}
\end{equation}
if the  domain is restricted to the positive magnetic spectral
subspace $\cH_{A+,2}:=\Lambda_{A+,2}(\cA(H_1({\Bbb R}^3)\otimes {\Bbb C}^4)^2)$ of the two electrons where $H_1({\Bbb R}^3)\otimes {\Bbb C}^4$ is the domain of $D_A^{(k)}.$
The projectors are defined by
$$\Lambda_{A+,2}\;=\;\Lambda_{A,+}^{(1)}\otimes \Lambda_{A,+}^{(2)},\qquad \Lambda_{A,+}^{(k)}:=\;\frac12\left( 1\;+\;\frac{D_A^{(k)}}{E_A^{(k)}}\right),$$
\begin{equation}\label{1.4a}
E_A^{(k)}:=\;|D_A^{(k)}|\;=\;\sqrt{(\bfp_k-e\bfA(\bfx_k))^2\,-\,e\bfsigma^{(k)} \bfB(\bfx_k)\,+m^2}\;\geq\;m
\end{equation}
and $\bfB = \bfnabla \times \bfA$ is the magnetic field. 
We note that the gauge invariance of the transformed operator
is preserved \cite{LSS1}.

If the field energy $E_f\,=\frac{1}{8\pi}\int_{{\Bbb R}^3} B^2(\bfx)\,d\bfx\;$ is added to $H_2$, positivity of the Brown-Ravenhall operator
can be established.
This relies on the condition $\|B\|<\infty$ to render $E_f$ finite, which excludes constant magnetic fields.
Lieb, Siedentop and Solovej \cite{LSS} have proven positivity (i.e. stability) for the $K$-nuclei $N$-electron Brown-Ravenhall operator, in case of $Z\leq 56.\;$ 
In the present context   $E_f$ can be disregarded, keeping in mind that it just leads to a global shift of the spectrum.

The aim of the present work is to provide the HVZ theorem for $H_2^{BR}$ which localizes the bottom of the essential spectrum \cite{Hu,Wi,Zhi}.
This theorem was proven for the multiparticle Brown-Ravenhall operator describing electrons in the
Coulomb potential of subcritical charge ($Z\leq 124$), but with $\bfA = {\bf 0}$ \cite{Jaku2,MV}. The method of proof,
which we also will adopt here, is based on the work of Lewis, Siedentop and Vugalter \cite{LSV} for the HVZ theorem concerning the scalar
pseudo-relativistic multiparticle (Herbst-type) Hamiltonian.

The main difference to the field-free case results from the kinetic energy being described by an {\it integral} operator (instead of a multiplication operator in momentum space).
In order to carry out the necessary computations an estimate of its kernel
is needed. If we restrict ourselves to constant magnetic fields we can profit from the relation to
 the heat kernel of the Schr\"{o}dinger operator which has been studied extensively (\cite{Avron,Simon,LT97,Erdos} and references therein).

Before we prove the HVZ theorem (in sections 5 and 6),

\begin{theorem}[HVZ theorem]\label{t1}
Let $H_2^{BR}$ be the two-electron  Brown-Ravenhall  operator from (\ref{1.3a}) where 
$\bfB$ is a constant magnetic field and where the Coulomb potential strength is restricted to $\gamma < \gamma_c = 0.629\;\;(Z\leq 86).$ Then its essential spectrum is given by
\begin{equation}\label{1.5a}
\sigma_{ess}(H_2^{BR})\;=\;[\Sigma_0,\infty)
\end{equation}
where $\Sigma_0$ is the ground state energy of the one-electron ion, increased by the rest mass of the second electron,
\end{theorem}

{\noindent we provide the relative form boundedness of the potential of $H_2^{BR}$ with respect to the kinetic energy operator} (which requires the restriction $\gamma <\gamma_c$; section 2).
For handling the IMS-type localization formula \cite[p.28]{CFKS}, entering into the proof of the HVZ theorem,
we found it convenient to work  in a representation of the Brown-Ravenhall operator which invokes the Foldy-Wouthuysen transformation $U_0$
instead of the projectors $\Lambda_{A,+}$ (section 3).
In section 4 we establish the necessary estimates for the kernels of $E_A$ and  $U_0$ as well
as for their commutators with some simple scaling functions.

We call an operator $\cO\;\;\frac{1}{R}$-bounded if $\cO$ is bounded by $c/R$ with some constant $c$ and $R\geq 1.$

\section{Relative form boundedness of the electric potential}
\setcounter{equation}{0}

For $\psi_+ \in \cH_{A+,2}$  we have $D_A^{(k)}\psi_+=E_A^{(k)}\psi_+.$
Therefore, $E_{A,tot}:= E_A^{(1)}+E_A^{(2)}$ can be identified with the kinetic energy operator.
We require $\bfA \in L_{2,loc}({\Bbb R}^3)$ \cite{Avron} which guarantees the essential self-adjointness of $E_A^{(k)},\;\;k=1,2 $ \cite{Jaku1}.
Moreover, we use the gauge $\bfnabla \cdot \bfA =0$.

\begin{lemma}\label{l1}
The Brown-Ravenhall operator $H_2^{BR}$ for $\bfA \in L_{2,loc}({\Bbb R}^3)$ is well-defined in the form sense on $\Lambda_{A+,2}(\cA(H_{1/2}({\Bbb R}^3) \otimes {\Bbb C}^4)^2)$ and is bounded from below for $\gamma <\gamma_c= 0.629.$
\end{lemma}

$H_2^{BR}$ thus extends to a self-adjoint operator for $\gamma<\gamma_c$ by means of the Friedrichs extension.

\begin{proof}
In order to show the relative form boundedness of the potential with respect to $E_{A,tot}$ let
 us assume that $B$ is bounded, $\|B\|_\infty := B_0 <\infty.$

For the two-particle potential we have the estimate \cite{BBHS} for $\psi \in \cA(H_{1/2}({\Bbb R}^3) \otimes {\Bbb C}^4)^2$, using Kato's inequality,
\begin{equation}\label{2.1}
(\psi,V^{(12)}\,\psi)\;\leq\; \frac{e^2\pi}{2}\;(\psi,p_1\,\psi) 
\;\leq\; \frac{e^2\pi}{2}\;(\psi,\sqrt{p_1^2+p_2^2+2m^2}\;\psi).
\end{equation}
Furthermore we employ the diamagnetic inequality (\cite{IK}, see also \cite{Avron} and references therein) which holds in arbitrary dimension.
Defining $S_A^{(k)}:= [(\bfp_k-e\bfA(\bfx_k))^2+m^2]^\frac12$,
 it can for $N$ particles be written in the following way  \cite{FLS,Avron},
$\,|(\sum_{k=1}^N S_A^{(k)2})^{-1/n}\,\psi|\leq(\sum_{k=1}^N p_k^2+Nm^2)^{-1/n}\;|\psi|,\;n=1,2,4,...\,.$
Upon multiplication with some function $f>0$ satisfying
$\|f(\bfx_1,...,\bfx_N)(\sum_{k=1}^N p_k^2+Nm^2)^{-1/n}\|\,\leq c_n$ we obtain
\begin{equation}\label{2.2c}
\left\| f\,\frac{1}{(\sum_{k=1}^N S_A^{(k)2})^{1/n}}\;\psi\right\|^2\;\leq\; \left\|f\;\frac{1}{(\sum_{k=1}^N p_k^2+Nm^2)^{1/n}}\,|\psi|\right\|^2
\;\leq\; c_n^2\;\|\psi\|^2.
\end{equation}
Let us choose $N=2,\;n=4,\;f(\bfx_1,\bfx_2)=e\,|\bfx_1-\bfx_2|^{-1/2}$ and $\psi =(\sum\limits_{k=1}^N S_A^{(k)2})^{1/4}\psi_+.\;$
Then (\ref{2.2c}) turns into
\begin{equation}\label{2.3c}
(\psi_+,V^{(12)}\,\psi_+)\;\leq\; \frac{e^2\pi}{2}\;(\psi_+,(\sum_{k=1}^2 S_A^{(k)2})^{1/2}\,\psi_+)\;
\leq\; \frac{e^2\pi}{2}\;(\psi_+,\sum_{k=1}^2 S_A^{(k)}\;\psi_+)
\end{equation}
with the constant from (\ref{2.1}). Using $S_A^{(k)2}\,=E_A^{(k)2}+e\bfsigma^{(k)}\bfB(\bfx_k)$ and
$|\bfsigma^{(k)}  \bfB(\bfx_k)|\,=B(\bfx_k)\leq B_0$ we estimate the r.h.s. of (\ref{2.3c}) further such that
\begin{equation}\label{2.3}
(\psi_+,V^{(12)}\,\psi_+)\;\leq\; 
\frac{e^2\pi}{2}\;(\psi_+,E_{A,tot}\,\psi_+) \;+\;e^2\pi\;(eB_0)^\frac12\;\|\psi_+\|^2.
\end{equation}
The relative $E_A^{(k)}$-form boundedness of the single-particle potential was shown in \cite{Jaku1},
\begin{equation}\label{2.4}
|(\psi_+,V^{(k)}\,\psi_+)|\;\leq\; \gamma \frac{\pi}{2}\;(\psi_+,E_A^{(k)}\,\psi_+)\;+\;\gamma\frac{\pi}{2}\;(eB_0)^\frac12\;\|\psi_+\|^2
\end{equation}
for $k=1,2.$ Therefore, we have
\begin{equation}\label{2.5}
|(\psi_+,(V^{(1)}+V^{(2)}+V^{(12)})\,\psi_+)|\,\leq \left( \frac{\gamma \pi}{2}\,+\,\frac{e^2\pi}{2}\right)(\psi_+,E_{A,tot}\,\psi_+)\,+
\, C_1(B_0)\,\|\psi_+\|^2,
\end{equation}
with $C_1(B_0)=\pi(\gamma+\,e^2)(eB_0)^{1/2}$ and form bound smaller than one for $\gamma < \gamma_c:= \frac{2}{\pi}\,-e^2 \approx 0.629.$ The lower bound $-C_1(B_0)$ of $H_2^{BR}$, derived from (\ref{2.5}), decreases with increasing magnetic field strength.
\end{proof}

\section{Alternative representation of $H_2^{BR}$ and the heat kernel of $E_A^2$}
\setcounter{equation}{0}

Following our strategy in the field-free case \cite{Jaku2} we use a representation of the
Brown-Ravenhall operator where its single-particle contributions do not depend on the coordinate of the second particle.
We note that any $\psi_+ \in \cH_{A+,2}$ can be represented in terms of a single-particle Foldy-Wouthuysen transformation $U_0$ \cite{DeV} (we will drop the superscript $(k)$ throughout when referring to the single-particle case),
$$U_0\;=\;A_E\left( 1\,+\,\beta \;\frac{\bfalpha(\bfp-e\bfA)}{E_A+m}\right),$$
\begin{equation}\label{7.1}
A_E\;=\;\left( \frac{E_A+m}{2E_A}\right)^\frac12.
\end{equation}
This operator has the advantage of being unitary and hence norm preserving (in  contrast to the projector $\Lambda_{A,+}$). We have \cite{Jaku1}
\begin{equation}\label{7.2}
\psi_+\;=\;(U_0^{(1)}U_0^{(2)})^{-1}\;\frac12(1+\beta^{(1)})\;\frac12 (1+\beta^{(2)})\;u
\end{equation}
where the inverse $(U_0)^{-1}$ follows from (\ref{7.1}) if $\beta$ is replaced by $-\beta$ and where $u \in \cA(H_1({\Bbb R}^3)\otimes {\Bbb C}^4)^2.\;$
Since $\frac12 (1+\beta^{(k)})$ projects onto the upper components of the spinor
associated with particle $k$ one may without restriction assume that the lower components of $u$ are zero (and omit $\frac12 (1+\beta^{(k)}),\;k=1,2).\;$
Thus, from (\ref{1.3a}),
$$(\psi_+, H_2^{BR}\,\psi_+)\;=\;(u,\, U_0^{(1)} U_0^{(2)} \left\{ \sum_{k=1}^2 (D_A^{(k)}+V^{(k)})\,+V^{(12)}\right\}\;(U_0^{(1)}U_0^{(2)})^{-1}\;u)$$
\begin{equation}\label{7.3}
=:\;(u,h_2^{BR}\;u).
\end{equation}
Using $U_0^{(k)}D_A^{(k)}\,(U_0^{(k)})^{-1}\,=\beta^{(k)}E_A^{(k)}$ \cite{DeV} the kinetic energy contribution to (\ref{7.3}) can be simplified to
\begin{equation}\label{7.4}
(u,\,U_0^{(1)}U_0^{(2)}\,D_A^{(k)}\,(U_0^{(1)}U_0^{(2)})^{-1}\;u)\;=\;(u,E_A^{(k)}\;u).
\end{equation}
Therefore we can identify
\begin{equation}\label{7.5a}
h_2^{BR}\;=\;\sum_{k=1}^2\left( E_A^{(k)}\,+\,U_0^{(k)}V^{(k)}\,(U_0^{(k)})^{-1}\right)\;+\;U_0^{(1)}U_0^{(2)}\,V^{(12)}\,(U_0^{(1)}U_0^{(2)})^{-1}
\end{equation}
keeping in mind that its quadratic form has to be taken with spinors of vanishing lower components.

\vspace{0.2cm}

The defining equation (\ref{1.4a}) for $E_A$  reveals the close connection between $E_A^2$ 
and the Schr\"{o}dinger operator  $H_s:=(\bfp-e\bfA)^2$ for a free  electron in a magnetic field.

Let us now restrict ourselves to a constant magnetic field, $\bfB=\bfB_0$,
and choose the $\bfe_3$-direction along $\bfB_0.$ Using the  gauge $\bfnabla \cdot \bfA=0$ we take \cite{Avron}
\begin{equation}\label{a.1} 
\bfA(\bfx)\;=\;\frac12\;(\bfB_0 \times \bfx)\;=\;\frac12\;B_0\;(-x_2,x_1,0).
\end{equation}
 Then the difference between the two operators $E_A^2$ and $H_s$ is a constant in space.
This allows us to adopt the properties of $H_s$.
For a $\bfB$-field with constant direction the Schr\"{o}dinger operator $H_s = p_3^2 \,+(p_1-eA_1)^2\,+(p_2-eA_2)^2$ separates, and the magnetic field problem reduces to two dimensions.

Let us denote by $\cO(\bfx,\bfx')$ for $\bfx,\bfx' \in {\Bbb R}^3$
the kernel of an integral operator $\cO$. For constant $\bfB=\bfB_0$ the (Mehler) heat kernel of $H_s$, i.e. the kernel of $e^{-tH_s}$, is known explicitly (see e.g. \cite{Avron},\cite[p.168]{Simon},\cite{LT97}),
$$e^{-tH_s}(\bfx,\bfx')\;=\;\frac{1}{(4\pi t)^{\frac12}}\;\frac{eB_0}{4\pi \sinh(eB_0t)}\;e^{-i\frac{e B_0}{2}(x_1x'_2-x_2x'_1)}$$
\begin{equation}\label{a.5}
\cdot e^{-(x_3-x'_3)^2/(4t)}\;e^{-\frac{eB_0}{4}\,\coth(eB_0t)\,[(x_1-x'_1)^2+(x_2-x'_2)^2]}
\end{equation}
where $t>0$ is the semigroup parameter\footnote{Note that the phase in (\ref{a.5}) differs
in sign from the one given by Loss and Thaller \cite{LT97} because they
consider $\tilde{H}_s=(\bfp+e\bfA)^2$ in place of $H_s$.}.
 Since $H_s$ is essentially self-adjoint on $C_0^\infty({\Bbb R}^3)$, its heat  kernel satisfies the symmetry property $e^{-tH_s}(\bfx,\bfx')\,=\\e^{-tH_s}(\bfx',\bfx)^\ast.$

The heat kernel of $E_A^2$ follows from
\begin{equation}\label{a.5a}
e^{-tE_A^2}(\bfx,\bfx')\;=\;e^{-tm^2}\;e^{te\bfsigmas\bfB_0}\;e^{-tH_s}(\bfx,\bfx').
\end{equation}
For the subsequent estimates we need a series of inequalities for the hyperbolic functions,
\begin{equation}\label{3.11}
z \coth z \;\leq\; 1+z \quad\mbox{ and} \quad \sinh z\;\geq\;z \quad \mbox{for } z\geq 0,
\end{equation}
$$z\,\coth z\;\geq\;1,\qquad \frac{z\,e^z}{\sinh z}\;\leq\;1+2z.$$
It is well known (see e.g. \cite[p.35]{Simon}) that, 
using (\ref{3.11}), $e^{-tH_s}(\bfx,\bfx')$ can be estimated from above by the heat kernel of the free Schr\"{o}dinger
operator,
\begin{equation}\label{a.6}
\left| e^{-tH_s}(\bfx,\bfx')\right|\;\leq\; \frac{1}{(4\pi t)^{\frac{3}{2}}}\;e^{-(\bfx-\bfx')^2/(4t)}\;=\;e^{-tp^2}(\bfx,\bfx').
\end{equation}
For (\ref{a.5a}), a less restrictive estimate of the prefactor is required, since
 $|e^{te\bfsigmas\bfB_0}|\,\leq\,e^{|te\bfsigmas\bfB_0|}\,=e^{teB_0}$. Using the last inequality of (\ref{3.11}) we have
\begin{equation}\label{a.6a}
\left| e^{-tE_A^2}(\bfx,\bfx')\right|\;\leq\; e^{-tm^2}\;\frac{1}{(4\pi t)^{\frac{3}{2}}}\;(1+2eB_0t)\;e^{-(\bfx-\bfx')^2/4t}
\end{equation}
and note in passing that for $m \neq 0$ the r.h.s. of (\ref{a.6a})
can  be further estimated by $c(B_0)\,e^{-tp^2}(\bfx,\bfx').$

\section{Estimate of the kernels of  $E_A$ and $U_0$ and  their commutators}
\setcounter{equation}{0}

Again we consider only the single-particle case and suppress the superscript $(k).$ We have

\begin{lemma}\label{l4}
For a magnetic field generated by the vector potential (\ref{a.1}) the kernel of the kinetic energy $E_A$ can be estimated by
\begin{equation}\label{b.4}
|E_A(\bfx,\bfx')|\;\leq\; \frac{C(B_0)}{|\bfx-\bfx'|^4}
\end{equation}
where the constant $C(B_0)$ increases quadratically with the field strength $B_0.$
We even have for some $0<\epsilon<m$,
\begin{equation}\label{b.4a}
|E_A(\bfx,\bfx')|\;\leq\; \frac{C(B_0)}{|\bfx-\bfx'|^4} \;e^{-(m-\epsilon)|\bfx-\bfx'|}.
\end{equation}
\end{lemma}

We remark that Frank, Lieb and Seiringer \cite{FLS} have derived a (\ref{b.4})-type estimate for  the scalar pseudorelativistic Hamiltonian  (where $E_A$ is replaced by $|\bfp-e\bfA|$ with $\bfA \in L_{2,loc}({\Bbb R}^3)$) in the context of the localization formula, with a constant $C$ independent of the magnetic field.

\begin{proof}
We use the integral representation \cite{Avron}
\begin{equation}\label{a.7}
\frac{1}{E_A}\;=\;\frac{1}{\sqrt{\pi}}\;\int_0^\infty \frac{dt}{\sqrt{t}}\;e^{-tE_A^2}
\end{equation}
to write
\begin{equation}\label{b.5}
E_A\;=\;\lim_{t \rightarrow 0}\;\frac{1-e^{-tE_A^2}}{tE_A}\;
=\;\lim_{t \rightarrow 0} \;\frac{1}{\sqrt{\pi}\,t}\,\int_0^\infty \frac{d\tau}{\sqrt{\tau}}\;\left( e^{-\tau E_A^2}\,-\;e^{-(t+\tau)E_A^2}\right).
\end{equation}
Since for $t>0,\;\;e^{-tE_A^2}(\bfx,\bfx')$ is analytic in $t$ (see (\ref{a.5a}) with (\ref{a.5})) we can expand $e^{-(t+\tau)E_A^2}(\bfx,\bfx')$ in powers of $t$ for small $t$.
Using the Taylor formula, $f(t+\tau)\,=f(\tau)+t f'(\tau)+O(t^2)\;$ we need the derivative
$$\frac{d}{d\tau}\,e^{-\tau E_A^2}(\bfx,\bfx')\;=\;e^{-\tau E_A^2}(\bfx,\bfx')\;\left\{ -eB_0 \coth(eB_0\tau)\,-\,\frac{1}{2\tau}\,-\,m^2\,\right.\,
+\,e\bfsigma\bfB_0 $$
\begin{equation}\label{b.7}
+\left.\,\frac{1}{4\tau^2}(x_3-x'_3)^2\,+\,\frac{(eB_0)^2}{4}\;\frac{1}{\sinh^2(eB_0\tau)}\;\left[ (x_1-x'_1)^2+(x_2-x'_2)^2\right]\right\}.
\end{equation}
An estimate of the derivative of the heat kernel of $E_A^2$ for a wide class of smooth vector potentials is provided by Ueki \cite{Ueki}.
However, due to the presence of the $e \bfsigma \bfB$-term in $E_A^2$  the given estimate increases exponentially for $t \rightarrow \infty$ and cannot be used in the present context.

In order to get an estimate for the kernel of $E_A$ we employ further the estimates  (\ref{3.11}) and (\ref{a.6a})
 (as well as the triangle inequality). This leads to
$$|E_A(\bfx,\bfx')|\;=\; \left| \frac{-1}{\sqrt{\pi}} \int_0^\infty
\frac{d\tau}{\sqrt{\tau}}\left( \frac{d}{d\tau}\;e^{-\tau E_A^2}(\bfx,\bfx')\right)\right|$$
$$\leq\;\frac{1}{8\pi^2}\int_0^\infty \frac{d\tau}{\tau^2}\;e^{-\tau m^2}\,e^{-(\bfx-\bfx')^2/(4\tau)}\;\left\{ \frac{3}{2\tau}\,+\,5eB_0\,+\,m^2\right.$$
\begin{equation}\label{b.8}
\left.+\;\frac{eB_0}{2\tau}\,(\bfx-\bfx')^2\,+\,\frac{1}{4\tau^2}\,(\bfx-\bfx')^2\,+\,(m^2+2eB_0)\,2eB_0\tau\right\}.
\end{equation}
Performing the integrals with the help of Appendix A, we get, setting $\bfz':=\bfx'-\bfx,$
$$|E_A(\bfx,\bfx')|\;\leq\; \frac{1}{2\pi^2}\left\{ K_1(mz')\,\left[ \frac{14m}{z^{'3}}\,+\,\frac{2m^3+7meB_0}{z'}
\right]\right.$$
\begin{equation}\label{b.9}
+\;\left.K_0(mz')\,\left[ \frac{7m^2}{z^{'2}}\,+\,2eB_0\,(m^2+eB_0)\right] \right\}.
\end{equation}
According to the behaviour of the modified Bessel functions $K_\nu$(see Appendix A),
the function in curly brackets diverges like $1/z^{'4}$ for $z'\rightarrow 0$ and decreases (for $m\neq 0$) like $e^{-mz'}/z^{'1/2}$ for $z' \rightarrow \infty$. Based on the continuity of this function for $z' \in {\Bbb R}_+$ one gets the estimate $|E_A(\bfx,\bfx')|\,\leq\frac{c_0}{|\bfx-\bfx'|^4}$ where $c_0$ increases quadratically with $B_0.$

From the less restrictive estimate $|E_A(\bfx,\bfx')|\,\leq \frac{c(B_0)}{z^{'4}}\,(1+z^{'7/2})\,e^{-mz'}\,$
one obtains an exponential decay
for $0<\epsilon <m$ (since $z^{'\nu}e^{-\epsilon z'}$ is bounded for any $\nu>0$), which proves (\ref{b.4a}).
\end{proof}

For the bounded operator $U_0$ the singularity of the kernel is weaker.

\begin{lemma}\label{l3s}
For a constant magnetic field and $0<\epsilon<m$ the kernel of the Foldy-Wouthuysen transformation $U_0$ can be estimated by
\begin{equation}\label{4.8s}
\left| U_0(\bfx,\bfx')\right|\;\leq\; \frac{c(B_0)}{|\bfx-\bfx'|^3}\;e^{-(m-\epsilon)|\bfx-\bfx'|},
\end{equation}
where $c(B_0)$ increases quadratically with the strength $B_0$ of the magnetic field.
\end{lemma}

\begin{proof}
From the definition (\ref{7.1}) we have $U_0=A_E+\,\frac{\beta}{\sqrt{2}}\,\frac{\bfalphas(\bfp-e\bfA)}{\sqrt{E_A(E_A+m)}}\;$
and we estimate the kernels of the two summands separately.

Concerning the second summand, (\ref{a.7}) leads to
\begin{equation}\label{4.9s}
\frac{1}{\sqrt{E_A}\sqrt{E_A+m}}\;=\;\frac{1}{\pi}\int_0^\infty \frac{dt'}{\sqrt{t'}}\int_0^\infty \frac{d\tau}{\sqrt{\tau}}\,e^{-\tau m}\,e^{-(\tau+t')E_A}.
\end{equation}
Abbreviating $t:= \tau+t'$ we make use of the integral
 representation \cite{FLS}
\begin{equation}\label{3.14b}
e^{-tE_A}\;=\;\frac{1}{\sqrt{\pi}}\int_0^\infty\frac{d\tau'}{\sqrt{\tau'}}\;e^{-\tau'}\;e^{-(t^2/4\tau')E_A^2}.
\end{equation}
Thus
\begin{equation}\label{4.11s}
\frac{\bfalpha(\bfp- e\bfA)}{\sqrt{E_A(E_A+m)}}(\bfx,\bfx')\;=\;
\frac{1}{\pi^{\frac{3}{2}}}\int_0^\infty \frac{dt'}{\sqrt{t'}}\int_0^\infty \frac{d\tau}{\sqrt{\tau}}\;e^{-\tau m}
\end{equation}
$$\cdot \int_0^\infty \frac{d\tau'}{\sqrt{\tau'}}\,e^{-\tau'}\,\left( -i\bfalpha \bfnabla_{\bfx}-\bfalpha e\bfA\right)\;e^{-(t^2/4\tau')E_A^2}(\bfx,\bfx').$$
For the derivative we have from (\ref{a.5a}) with (\ref{a.5})
$$-i \bfalpha \bfnabla_{\bfx} e^{-tH_s}(\bfx,\bfx')\;=\;\left\{ \frac{i}{2t}\,\alpha_3\,(x_3-x'_3)\;+\;i\,\frac{eB_0}{2}\,\coth(eB_0t)\;\right.$$
\begin{equation}\label{a.12}
\cdot [\alpha_1 (x_1-x'_1)+\alpha_2(x_2-x'_2)]
\left.-\;\frac{eB_0}{2}\;[\alpha_1x'_2-\alpha_2x'_1]\right\}\;e^{-tH_s}(\bfx,\bfx')
\end{equation}
and further
\begin{equation}\label{a.11}
-\bfalpha e\bfA\;e^{-tH_s}(\bfx,\bfx')\;=\;\frac{eB_0}{2}\;(\alpha_1x_2-\alpha_2 x_1)\;e^{-tH_s}(\bfx,\bfx')
\end{equation}
which have to be inserted into (\ref{4.11s}).

In the following  we make use of some relations.
From $\alpha_i^2=1,\; \sigma_i^2=1,\;\;i=1,2,3,$ and $\alpha_i \alpha_k = - \alpha_k\alpha_i$ for $i\neq k$ 
 we have 
$|\alpha_1(x_1-x'_1)+\alpha_2(x_2-x'_2)|\,=\sqrt{(x_1-x'_1)^2+(x_2-x'_2)^2}$. Moreover, employing $|\beta|=1$, (\ref{3.11}) and (\ref{a.6a}) we obtain
$$\left| \frac{\bfalpha( \bfp- e\bfA)}{\sqrt{E_A(E_A+m)}}(\bfx,\bfx')\right|\;\leq\; \frac{1}{\pi^3}\;|\bfx-\bfx'|\int_0^\infty \frac{dt'}{\sqrt{t'}}
\int_0^\infty \frac{d\tau}{\sqrt{\tau}}\,e^{-\tau m}\,\frac{1}{t^3}$$
\begin{equation}\label{4.14s}
\cdot \int_0^\infty d\tau'\,e^{-\tau'}\,e^{-(\bfx-\bfx')^2\tau'/t^2}
\,e^{-t^2m^2/4\tau'}\;\left\{ \frac{4\tau^{'2}}{t^2}\,+\,3eB_0\,\tau'\,+\,\frac{(eB_0)^2}{2}\,t^2\right\}.
\end{equation}
The $d\tau'$-integral can be evaluated with the help of (\ref{A.6a}), resulting in
\begin{equation}\label{4.15s}
\int_0^\infty d\tau' \cdots \;=\;t^4\left\{ m^3\,\frac{K_3(m\xi)}{\xi^3}\,+\,\frac{3}{2}\,eB_0m^2\,\frac{K_2(m\xi)}{\xi^2}\,+\,\frac{(eB_0)^2}{2}\,m\,\frac{K_1(m\xi)}{\xi}\right\}
\end{equation}
where $\xi:= \sqrt{t^2+(\bfx-\bfx')^2}.$
For the further estimate we note that the modified Bessel functions $K_\nu$ are monotonously decreasing in $(0,\infty)$, as are the inverse powers of $\xi$.
Therefore, (\ref{4.15s}) is estimated from above if $\xi$ is replaced by $y:= \sqrt{\tau^2+t^{'2}+(\bfx-\bfx')^2}\,\leq \xi.$
With the additional estimate $e^{-\tau m}\leq 1$
the integrand is symmetric in $\tau$ and $t'$ such that $t^4/t^3=\tau+t'$ can be
replaced by $2t'$. With these manipulations the $d\tau$-integral can be evaluated analytically by  (\ref{A.6b}). The result is,
defining $a^2:=t^{'2}+(\bfx-\bfx')^2,$
$$\left| \frac{\bfalpha( \bfp- e \bfA)}{\sqrt{E_A(E_A+m)}}(\bfx,\bfx')\right|\;\leq\; \frac{2}{\pi^3}\;|\bfx-\bfx'|\int_0^\infty dt'\,\sqrt{t'}\,\frac{2^{-\frac{3}{4}}\Gamma(\frac{1}{4})}{m^{\frac{1}{4}}}$$
\begin{equation}\label{4.16s}
\cdot \left\{ m^3\,\frac{K_{\frac{11}{4}}(ma)}{a^{\frac{11}{4}}}\,+\,\frac{3}{2}\,eB_0m^2\,\frac{K_{\frac{7}{4}}(ma)}{a^{\frac{7}{4}}}\,+\,\frac{(eB_0)^2}{2}\,m\,\frac{K_{\frac{3}{4}}(ma)}{a^{\frac{3}{4}}}
\right\}.
\end{equation}
Applying again (\ref{A.6b}) for the remaining integral turns the r.h.s. of (\ref{4.16s}) into
$$\frac{\sqrt{2}}{\pi^2}\,\left\{ m^2\,\frac{K_2(m|\bfx-\bfx'|)}{|\bfx-\bfx'|}\,+\,\frac{3}{2}\,eB_0m\,K_1(m|\bfx-\bfx'|)\right.$$
\begin{equation}\label{4.17s}
\left.+\;\frac{(eB_0)^2}{2}\,|\bfx-\bfx'|\,K_0(m|\bfx-\bfx'|)\right\}.
\end{equation}
Following the argumentation below (\ref{b.9}) one obtains the result\\
$| \frac{\bfalphas(\bfp-e\bfA)}{\sqrt{E_A(E_A+m)}}(\bfx,\bfx')|\,\leq\,\frac{c_1(B_0)}{|\bfx-\bfx'|^3}\,e^{-(m-\epsilon)|\bfx-\bfx'|}.\;$
It is shown in Appendix C that the kernel of $A_E$ obeys the same 
estimate. Thus
\begin{equation}\label{4.18s}
|U_0(\bfx,\bfx')|\;\leq\; \frac{1}{|\bfx-\bfx'|^3}\,e^{-(m-\epsilon)|\bfx-\bfx'|}\,\left[ \tilde{c}(B_0)\,+\,\|\frac{\beta}{\sqrt{2}}\|\,c_1(B_0)\right]
\end{equation}
which completes the proof.
\end{proof}

We remark that with the tools from the proof of Lemma \ref{l3s} 
it is straightforward to show that a (\ref{4.8s})-type estimate holds also for the
kernel of the projector,  $|\Lambda_{A,+}(\bfx,\bfx')|\,\leq \frac12 \delta(\bfx-\bfx')
+ \frac{C(B_0)}{|\bfx-\bfx'|^3}e^{-(m-\epsilon)|\bfx-\bfx'|}$.

\vspace{0.2cm}

Let us now turn to the estimates of the commutators which are needed in the context of the proof of the HVZ theorem.
In part they are based on a lemma (with its corollary), proven in
 \cite[section 8]{Jaku2}.

\begin{lemma}\label{l7a}
Let $\cO$ be a single-particle operator the kernel of which satisfies \\ $|\cO(\bfx,\bfx')| \leq\,\frac{c_0}{|\bfx-\bfx'|^3}\;$
with some constant $c_0.$
Let $g_0 \in C^\infty({\Bbb R}^3)$ (or $g_0 \in C({\Bbb R}^3) \cap C^1({\Bbb R}^3 \backslash \{0\})$ and $g_0(0)=0)$ be a real  function of $\bfx$ and let its derivative be bounded.
Then for $\varphi \in C_0^\infty({\Bbb R}^3)\otimes {\Bbb C}^4$ and $\psi \in L_2({\Bbb R}^3) \otimes {\Bbb C}^4$ one has
\begin{equation}\label{7.14}
|(\psi,\,[\cO,g_0]\;\frac{1}{x}\;\varphi)|\;\leq\; c\;\|\psi\|\;\|\varphi\|
\end{equation}
with a constant $c$ (depending on $c_0$) where $[\cO,g_0]:= \cO g_0-g_0 \cO.\;$

If $g$ is a function of $\bfx/R$, where $R>1$ is a scaling parameter, but otherwise with the same properties as $g_0$ then
\begin{equation}\label{7.15}
|(\psi,\,[\cO,g]\;\frac{1}{x}\;\varphi)|\;\leq\; \frac{c}{R}\;\|\psi\|\;\|\varphi\|.
\end{equation}
\end{lemma}

\begin{corollary}\label{c1i}
Let $g$ and $g_0$ be as in Lemma \ref{l7a} with $\bfx:= \bfx_1-\bfx_2$ or $\bfx=(\bfx_1,\bfx_2) \in {\Bbb R}^6$ (i.e. $g,g_0 \in C^\infty({\Bbb R}^6)).\;$
Then one has
\begin{equation}\label{4.i1}
(\psi, \,\frac{1}{|\bfx_1-\bfx_2|}\;[\cO,g_0]\,\varphi)|\;\leq\;c\;\|\psi\|\;\|\varphi\|
\end{equation}
and
\begin{equation}\label{4.i2}
|(\psi,\,\frac{1}{|\bfx_1-\bfx_2|}\,[\cO,g]\,\varphi)|\;\leq\; \frac{c}{R}\;\|\psi\|\;\|\varphi\|.
\end{equation}
\end{corollary}

{\noindent  Note that an operator and its adjoint have the same bound.}

Since according to Lemma \ref{l3s} the operator $U_0$ satisfies 
$|U_0(\bfx,\bfx')|\,\leq\,\frac{c(B_0)}{|\bfx-\bfx'|^3},$ we can apply Lemma \ref{l7a} with a suitable function $g$ to obtain the
 commutator estimate (with a generic constant $c$),
\begin{equation}\label{4.3i}
|(\psi,\,[U_0,g]\,\frac{1}{x}\,\varphi)|\;\leq\; \frac{c(B_0)}{R}\;\|\psi\|\;\|\varphi\|.
\end{equation}

We note that the same estimate is valid if $U_0$ is replaced by its inverse $(U_0)^{-1}.$
Also   the other three estimates, (\ref{7.14}), (\ref{4.i1}) and (\ref{4.i2})  hold for the operators $U_0$ and $(U_0)^{-1}$. 

It is more involved to obtain a commutator estimate for the kinetic energy
operator, because the singularity of its kernel is of the fourth power of $|\bfx-\bfx'|^{-1}.$

\begin{lemma}\label{l5}
Let $g \in C^\infty({\Bbb R}^3)$ be a real  function of $\bfx/R$ with bounded first and second derivative and  a  scaling parameter $R>1$. Then its commutator with the kinetic energy operator $E_A$ (for a constant magnetic field)  satisfies
\begin{equation}\label{4.11}
\|[E_A,g]\,\varphi\|\;\leq\; \frac{c(B_0)}{R}
\end{equation}
for $\varphi \in C_0^\infty({\Bbb R}^3) \otimes {\Bbb C}^4,\;\;\|\varphi\|=1,\,$
where the constant $c$ depends on the field strength $B_0$.
\end{lemma}

\begin{proof}
We have
\begin{equation}\label{4.12}
(\,[E_A,g]\,\varphi)(\bfx)\;=\;\int_{{\Bbb R}^3} d\bfx'\;E_A(\bfx,\bfx')\left\{ g\left(\frac{\bfx'}{R}\right) \,-\,g\left( \frac{\bfx}{R}\right) \right\}\;\varphi(\bfx').
\end{equation}
Let us define $\omega_1:= \frac{eB_0}{2}\,x_2,\;\;\omega_2:=-\frac{eB_0}{2}\,x_1,\;\;\omega_3:=0\;$ and $E_A^{\omega=0}(\bfx'-\bfx)$ by means of 
$E_A(\bfx,\bfx')=E_A^{\omega=0}(\bfx'-\bfx) e^{i\bfomegas(\bfx'-\bfx)}$ where 
$E_A^{\omega=0}$ is an even function and where we have used that $\bfomega \bfx' = \bfomega(\bfx'-\bfx).$

We aim at isolating the leading singularity of $E_A^{\omega=0}$ at $\bfx'=\bfx.$ With
\begin{equation}\label{4.13a}
E_A^{\omega=0}(\bfx'-\bfx)\;=\;- \frac{1}{\sqrt{\pi}} \int_0^\infty \frac{d\tau}{\sqrt{\tau}}\;\left(\frac{d}{d\tau} e^{-\tau E_A^{2(\omega=0)}}\!\!(\bfx'-\bfx)\right)
\end{equation}
and (\ref{b.7}) the integrand is analytic in $\bfx'-\bfx$ and in $\tau$ for $\tau >0.$
Therefore, the singular behaviour of $E_A^{\omega=0}$ in $\bfx'-\bfx={\bf 0}$
(which results from the factor $e^{-(\bfx'-\bfx)^2/4\tau}/\tau^n,\;\;n>1$)
 is found from an expansion of the integrand near $\tau =0$.
Setting $\bfz':=\bfx'-\bfx$ one obtains for the leading term $-\frac{1}{8\pi^2 \tau^2}\,e^{-z^{'2}/4\tau}\left( \frac{z^{'2}}{4\tau^2}-\frac{3}{2\tau}\right).\;$ Performing the $\tau$-integral over a small interval $(0,\epsilon)$ one thus gets
\begin{equation}\label{4.14a}
E_A^{\omega=0}(\bfz')\;=\;\frac{-1}{\pi^2 z^{'4}}\;+\;O(\frac{1}{z^{'2}}) \qquad \mbox{ for } z' \rightarrow 0.
\end{equation}
Therefore we decompose
\begin{equation}\label{4.15a}
E_A(\bfz')\;=\;\left( E_A^{\omega=0}(\bfz')\,+\,\frac{e^{-mz'}}{\pi^2 z^{'4}}\right)\;e^{i\bfomegas\bfz'}\;-\;\frac{e^{-mz'}}{\pi^2 z^{'4}}\;
e^{i\bfomegas\bfz'}.
\end{equation}

In order to check the boundedness of an operator $\cO$ the kernel of which can be estimated by a positive function $k$, viz. $|\cO(\bfx,\bfx')|\,\leq k(\bfx,\bfx'),$
the Lieb and Yau formula \cite{LY} can be applied. This formula is related  to the
Schur test for the boundedness of integral operators and tells us that $\cO$ is bounded if the following estimates hold for $k$,
\begin{eqnarray}\label{3.16}
I(\bfx):=& \int\limits_{{\Bbb R}^3} d\bfx'\;k(\bfx,\bfx')\;\frac{f(x)}{f(x')}&\leq\;C \nonumber\\
&&\nonumber\\
J(\bfx'):=& \int\limits_{{\Bbb R}^3} d\bfx\;k(\bfx,\bfx')\;\frac{f(x')}{f(x)}&\leq\;C
\end{eqnarray}
for all $x \in {\Bbb R}^3$ and $\bfx' \in {\Bbb R}^3$, respectively, where $f>0$ for $x>0$ is a suitable convergence generating function
and $C$ some constant. 

We will apply this formula to
the contribution to (\ref{4.12}) which arises from insertion of the first term of (\ref{4.15a}). 
For the function $g \in C^\infty({\Bbb R}^3)$ we can apply the mean value theorem to get
\begin{equation}\label{3.15}
\left| g(\frac{\bfx'}{R})\,-\,g(\frac{\bfx}{R})\right|\;=\;\left| \frac{\bfx-\bfx'}{R}\right|\;\left| (\bfnabla_{\!\!\frac{\bfx}{R}}g)(\frac{\bfxi}{R})\right|\;\leq\; \frac{c_0}{R}\;|\bfx-\bfx'|
\end{equation}
with some constant $c_0$ for $\bfxi$ on the line between $\bfx$ and $\bfx'$.
Then we obtain for the integral $I$  (which is identical to $J$ for the choice $f=1$),
$$I(\bfx)\;=\;\int_{{\Bbb R}^3} d\bfx'\left| \left( E_A^{\omega=0}(\bfz')\,+\,\frac{e^{-mz'}}{\pi^2z^{'4}}\right) e^{i\bfomegas \bfz'}\right|\;\left| g\left( \frac{\bfx'}{R}\right)-\,g\left( \frac{\bfx}{R}\right)\right|$$
\begin{equation}\label{4.16a}
\leq\;\int_{{\Bbb R}^3} d\bfz'\left| E_A^{\omega=0}(\bfz')\,+\,\frac{e^{-mz'}}{\pi^2z^{'4}}\right|\;\frac{c_0}{R}\,z'\;\leq\;\frac{c(B_0)}{R}.
\end{equation}
Since $|E_A^{\omega=0}(\bfz')\,+\frac{e^{-mz'}}{\pi^2z^{'4}}|\;$
behaves like $\frac{1}{z^{'3}}$ near $z'=0$, is analytic for $z'>0$ and decays exponentially at infinity (see (\ref{b.4a})), the  integral $I(\bfx)$ is finite and its bound is independent of $\bfx$. 

In order to treat the second contribution to $E_A(\bfz')$ we need the Taylor expansion of $g$ up
to second order. We use
the notation $g(\bfx/R)=:g_R(\bfx).$ Then
\begin{equation}\label{4.13}
g_R(\bfx') \,-\,g_R(\bfx)\;=\;(\bfx'-\bfx)\;\bfnabla g_R(\bfx)\;+\;\frac12 \sum_{k,l=1}^3 (x'_k-x_k)(x'_l-x_l)\;\nabla_k\nabla_l\,g_R(\bfxi)
\end{equation}
with $\bfxi$ on the line between $\bfx$ and $\bfx'$.  

The insertion of  the second term of (\ref{4.15a}) together with the second-order term of (\ref{4.13})  into (\ref{4.12}) provides the kernel of a bounded operator
(with bound $\frac{c}{R^2}$) according to the Schur test
  (again with $f=1$). The argumentation is the same as given in the context of (\ref{4.16a}) supplemented by the $1/R^2$-boundedness of the second derivatives of $g$.
Therefore it is sufficient to consider only the first term of (\ref{4.13}) in the remaining proof of the $1/R$-boundedness of $[E_A,g]\varphi$.

For $\omega=0$ denote the corresponding kernel by
\begin{equation}\label{4.15}
k_{0}(\bfz'):=  \;\frac{\bfc_1}{R} \;\frac{e^{-mz'}}{z^{'4}}\;\bfz'
\end{equation}
where $\bfc_1/R:=\bfnabla g_R(\bfx)/\pi^2$ is bounded and independent of $\bfz'$.
The kernel $k_0$  defines a bounded operator according to Stein's theorem \cite[\S2.3.2]{S}, \cite{MV}. Its requirements for the kernel (which should depend only
on one variable) are
\begin{eqnarray}\label{4.16}
&(i)& |k_0(\bfz')|\;\leq\; \frac{c}{z^{'3}} \nonumber\\
&(ii)& |\bfnabla k_0(\bfz')|\;\leq\; \frac{c}{z^{'4}}\\
&(iii)& \int\limits_{R_1<z'<R_2}k_0(\bfz')\;d\bfz'\;=\;0\qquad \mbox{ for } \;0<R_1<R_2<\infty.\nonumber
\end{eqnarray}
The conditions $(i)$ and $(iii)$ are obviously fulfilled, the latter since
$k_0$ is an odd function. For the proof of $(ii)$ we use that $me^{-mz'}/z^{'3}\,=m(z'e^{-mz'})/z^{'4}\leq c/z^{'4}.$

\vspace{0.2cm}
For the case  $\omega \neq 0$ we recall that proving (\ref{4.11}) is equivalent to showing that
$\,|(\psi, [E_A,g]\,\varphi)|\,\leq\frac{c(B_0)}{R}\,\|\psi\|\,$ for $\psi \in L_2({\Bbb R}^3)\otimes {\Bbb C}^4$ \cite[p.260]{RS1}.
Thus we have to consider
\begin{equation}\label{4.20}
\int_{{\Bbb R}^3} d\bfx\;\overline{\psi}(\bfx)\int_{{\Bbb R}^3} d\bfz'\;k_0(\bfz')\;e^{i\bfomegas \bfz'}\;\varphi(\bfx+\bfz').
\end{equation}
Since $\varphi $ is a $C_0^\infty$-function one has the Taylor formula which can be used to any order in the proof below. Its simplest form agrees with the mean value theorem,
\begin{equation}\label{4.21}
\varphi(\bfx+\bfz')\;=\;\varphi(\bfx)\,+\,\bfz'\;(\bfnabla\varphi)(\bfx+\lambda \bfz')
\end{equation}
where $\lambda \in (0,1).\;$

Insertion of the second contribution in (\ref{4.21}) into (\ref{4.20}) leads to
\begin{equation}\label{4.22}
S_0:=\;\frac{1}{R}\sum_{k,l=1}^3 c_{1k} \int_{{\Bbb R}^3} d\bfx \;\overline{\psi}(\bfx)\int_{{\Bbb R}^3} d\bfz'\;\frac{e^{-mz'}}{z^{'4}}\;z'_kz'_l\;e^{i\bfomegas \bfz'}\;(\bfnabla\varphi)_l(\bfx+\lambda \bfz').
\end{equation}
We estimate this integral in the following way (and subsequently apply the Schwarz inequality),
$$|S_0|\;\leq\; \frac{1}{R}\sum_{k,l=1}^3 |c_{1k}| \int_{{\Bbb R}^6}d\bfx \; d\bfz'\left( |\psi(\bfx)|\; \frac{e^{-\frac{mz'}{2}}}{z'}\right)\left( |(\bfnabla\varphi)_l(\bfx+\lambda \bfz')|\; \frac{e^{-\frac{mz'}{2}}}{z'}\right)$$
\begin{equation}\label{4.23a}
\leq\;\frac{1}{R} \sum_{k,l=1}^3 |c_{1k}|\left( \int_{{\Bbb R}^3}d\bfx \,|\psi(\bfx)|^2\int_{{\Bbb R}^3}d\bfz'\,\frac{e^{-mz'}}{z^{'2}}\right)^\frac12
\end{equation}
$$\cdot \left( \int_{{\Bbb R}^3} d\bfz' \,\frac{e^{-mz'}}{z^{'2}}\int_{{\Bbb R}^3}d\bfx\,| (\bfnabla \varphi)_l(\bfx+\lambda \bfz')|^2\right)^\frac12.$$
In the second $\bfx$-integral we make the variable transform
$\bfy:=\bfx+\lambda \bfz'$ and have
$\int_{{\Bbb R}^3} d\bfy\,|(\bfnabla \varphi)_l(\bfy)|^2\,\leq c_0\;$
since the derivative $\bfnabla \varphi$ is a $C_0^\infty$-function too (and hence in $L_2$).
Then both $\bfz'$-integrals are equal and finite, which proves $|S_0|\,\leq \frac{c}{R}\,\|\psi\|.$

For the first contribution in (\ref{4.21}) we have
\begin{equation}\label{4.23b}
\frac{1}{R}\sum_{k=1}^3 c_{1k} \int_{{\Bbb R}^3} d\bfx\;\overline{\psi}(\bfx)\;\varphi(\bfx)\int_{{\Bbb R}^3}d\bfz'\;\frac{e^{-mz'}}{z^{'4}}\;z'_ke^{i\bfomegas\bfz'}.
\end{equation}
The second integral can be evaluated analytically.
Using the symmetry of the integration intervals, one gets zero for $k=3.$ For $k=1,2$
we make the substitution $\bfxi:=\omega \bfz'$ with $\omega=\sqrt{\omega_1^2+\omega_2^2}$ and use spherical coordinates. Then (for $k=1$)
$$S_1:=\; \int_{{\Bbb R}^3} d\bfz'\;\frac{e^{-mz'}}{z^{'4}}\;z_1\;e^{i\bfomegas\bfz'}\;=\;\int_0^\infty \frac{d\xi}{\xi^2}\;e^{-\frac{m}{\omega}\xi}\int_0^\pi \sin \theta\;d\theta$$
\begin{equation}\label{4.41}
\cdot\int_0^{2\pi}d\varphi\;\xi\,\sin \theta\,\cos \varphi\;e^{i\xi \sin \theta\,(\frac{\omega_1}{\omega}\cos \varphi+\frac{\omega_2}{\omega}\sin \varphi)}.
\end{equation}
Since $-1\leq \omega_k/\omega\leq 1,\;$ we introduce the angle $\alpha$ by means of $\sin \alpha=\omega_1/\omega,\;\;\cos \alpha=\omega_2/\omega$
such that the phase reduces to $\xi \sin \theta \,\sin(\varphi + \alpha).$
Then we shift $\varphi$ as well as the integration interval by $-\alpha$ and obtain \cite[p.400,401]{G}
$$\int_0^{2\pi} d\varphi\;\cos \varphi\;e^{i\xi \sin \theta \,\sin(\varphi + \alpha)}\;=\;2i\int_0^\pi d\varphi'\;(\cos \varphi'\cos \alpha+\sin \varphi' \sin \alpha)$$
\begin{equation}\label{4.42}
\cdot \sin(\xi \sin\theta \sin \varphi')\;=\;2\pi i\;\frac{\omega_1}{\omega}\;J_1(\xi \sin \theta)
\end{equation}
where $J_\nu$ is a Bessel function of the first kind.

For $k=2$ we have $\sin \varphi \,=\sin (\varphi' -\alpha)\,=\sin \varphi' \cos \alpha -\cos \varphi' \sin \alpha\;$
in place of $\cos \varphi$ in (\ref{4.42}).
The result is $2\pi i\,\frac{\omega_2}{\omega}\,J_1(\xi \sin \theta).\;$
Using symmetry, we get for the second angular integral in (\ref{4.41}) \cite[p.740]{G}
\begin{equation}\label{4.43}
2\int_0^{\pi/2}\sin^2 \theta\;d\theta\;J_1(\xi \sin \theta)\;=\;\sqrt{\frac{2\pi}{\xi}}\;J_{3/2}(\xi).
\end{equation}
Thus \cite[p.711]{G}
$$S_k\;=\;(2\pi)^{3/2}i\;\frac{\omega_k}{\omega}\int_0^\infty \frac{d\xi}{\xi^{3/2}}\;e^{-\frac{m}{\omega}\xi}\;J_{3/2}(\xi)$$
\begin{equation}\label{4.44}
=\;\frac{4\pi i}{3}\;\frac{\omega_k}{\sqrt{m^2+\omega^2}}\;\;_2F_1(\frac12,\frac{3}{2},\frac{5}{2};\,\frac{\omega^2}{m^2+\omega^2})
\end{equation}
which is finite for all $\omega\neq 0.$
We note that for $m=0$ (or equivalenly, $\omega \rightarrow \infty$) one gets directly \cite[p.684]{G}
\begin{equation}\label{4.45}
\int_0^\infty \frac{d\xi}{\xi^{3/2}}\;J_{3/2}(\xi)\;=\;\frac{\sqrt{\pi}}{2\sqrt{2}}
\end{equation}
and hence $S_k(m=0)\,=\pi^2i\,\frac{\omega_k}{\omega}$
(which agrees with (\ref{4.44}) at $m=0$ \cite[p.1042]{G}).
Thus (\ref{4.23b}) can be estimated by $\frac{c}{R}\,\|\psi\|\;\|\varphi\|\;$ which completes the proof.
\end{proof}

\section{Proof of Theorem \ref{t1} (easy part)}

\setcounter{equation}{0}

Let us denote with $j=1,2$ the two electrons and with $j=0$ the nucleus
which generates the electric field. We define the two-cluster decompositions of $h_2^{BR}$,
\begin{equation}\label{5.2}
h_2^{BR}\;=\;E_{A,tot}\,+\,a_j\,+\,r_j,\qquad j=0,1,2,
\end{equation}
where $E_{A,tot}=E_A^{(1)}+E_A^{(2)}$ is the kinetic energy operator and $a_j$ collects the potential terms not involving particle $j$.
These three decompositions represent the cases where either one electron has moved to infinity ($j=1,2$) or where
both electrons have moved far away from the nucleus ($j=0$).
Then the residual interaction ($r_j$) is expected to tend to zero.
 For example, for $j=1$,
\begin{equation}\label{5.3}
a_1\;=\;U_0^{(2)}V^{(2)} \,(U_0^{(2)})^{-1},\quad r_1\;=\;
U_0^{(1)}V^{(1)}\,(U_0^{(1)})^{-1}\,+\,U_0^{(1)}U_0^{(2)}V^{(12)}(U_0^{(1)}U_0^{(2)})^{-1}.
\end{equation}
This allows us to define $\Sigma_0$ by means of
\begin{equation}\label{5.3a}
\Sigma_0:=\;\min_j\,\inf\,\sigma(E_{A,tot}+a_j).
\end{equation}

The proof of the HVZ theorem consists of two parts, conventionally called the 'easy part', $[\Sigma_0,\infty) \subset \sigma_{ess}(H_2^{BR})=\sigma_{ess}(h_2^{BR}),$
and the 'hard part',
 $\sigma_{ess}(h_2^{BR}) \subset [\Sigma_0,\infty)$.
Following Morozov and Vugalter \cite{MV} we shall work in coordinate space only.

For the easy part we use the strategy of Weyl sequences \cite{CFKS,Jaku2}.
Let $\lambda \in [\Sigma_0,\infty).$ Without restriction we can assume that $\Sigma_0 = \inf \sigma(E_{A,tot}+a_j)$ for $j=1.$
This relies on the symmetry of  the operator under electron exchange, while for $j=0$ we have $a_0=U_0^{(1)}U_0^{(2)}V^{(12)}(U_0^{(1)}U_0^{(2)})^{-1}\geq 0$ in contrast to $a_1\leq 0$ such that $\inf \sigma(E_{A,tot}+a_0) \geq \inf \sigma(E_{A,tot}+a_1).$
Since $E_{A,tot}+a_1$ does not contain any electron-electron
interaction it can be written as a sum of operators acting on different particles. This leads to a decomposition of the spectrum,
\begin{equation}\label{5.18}
\sigma(E_{A,tot}+a_1)\;=
\;\sigma(E_A^{(1)})\;+\;\sigma(E_A^{(2)}\,+\,U_0^{(2)}V^{(2)}(U_0^{(2)})^{-1}).
\end{equation}
The spectrum of $E_A^{(1)}$ is continuous and extends to infinity since the magnetic field does not
affect the electronic motion along $\bfB_0$ (which is the $\bfe_3$-direction). Therefore, $\sigma(E_{A,tot}+a_1)$ is also continuous such that $\lambda \in \sigma(E_{A,tot}+a_1)$ with the decomposition  $\lambda=\lambda_1+\lambda_2$ according to  (\ref{5.18}).

Let $(\varphi_n^{(1)})_\nin$ be a Weyl sequence for $\lambda_1$ consisting of normalized functions with $\varphi_n^{(1)} \stackrel{w}{\rightharpoonup} 0$ and
\begin{equation}\label{6.1}
\|(E_A^{(1)}-\lambda_1)\,\varphi_n^{(1)}\|\;\rightarrow \;0\qquad \mbox{ as } \ninfi.
\end{equation}
According to Appendix B (Lemma \ref{l6a}, which also holds for single-particle operators of the form $E_A^{(k)}+w^{(k)}$ )
we can in addition assume that $\varphi_n^{(1)} \in C_0^\infty({\Bbb R}^3 \backslash B_n(0))\otimes {\Bbb C}^4$.
Let $(\phi_n^{(2)})_\nin$ be a defining sequence for $\lambda_2$
with $\phi_n^{(2)} \in C_0^\infty({\Bbb R}^3) \otimes {\Bbb C}^4$ satisfying
$\|\phi_n^{(2)}\|=1$ and $\,\|(E_A^{(2)}+U_0^{(2)}V^{(2)}(U_0^{(2)})^{-1}\,-\lambda_2)\phi_n^{(2)}\|\,\rightarrow 0$ as $\ninfi.\;$
This implies, for a given $\epsilon >0,$ the existence of $N \in {\Bbb N}$ such that
\begin{equation}\label{6.2}
\|(E_A^{(2)}+U_0^{(2)}V^{(2)}(U_0^{(2)})^{-1}\,-\lambda_2)\;\phi_N^{(2)}\|\;<\;\epsilon.
\end{equation}
We claim that a subsequence of the sequence $(\cA\psi_n)_\nin$ with $\psi_n:= \varphi_n^{(1)} \phi_N^{(2)}$ is a Weyl sequence for $\lambda $
obeying $\,\|(h_2^{BR}-\lambda)\,\cA\psi_n\|\,\rightarrow 0$ for $\ninfi$ such that $\lambda \in \sigma_{ess}(h_2^{BR}).$
Disregarding for the moment the antisymmetrization, we estimate
\begin{equation}\label{6.3}
\|(h_2^{BR}-\lambda)\;\varphi_n^{(1)}\phi_N^{(2)}\|\;\leq\;\|(E_{A,tot}+a_1-\lambda)\;\varphi_n^{(1)}\phi_N^{(2)}\|\;
+\;\|r_1\,\varphi_n^{(1)}\phi_N^{(2)}\|.
\end{equation}
For the first term we have
$$\|(E_A^{(1)}-\lambda_1\, + E_A^{(2)} +U_0^{(2)}V^{(2)}(U_0^{(2)})^{-1}-\lambda_2)\;\varphi_n^{(1)}\phi_N^{(2)}\|$$
\begin{equation}\label{5.22a}
\leq\; \|\phi_N^{(2)}(E_A^{(1)}-\lambda_1)\,\varphi_n^{(1)}\|\,+\,\|\varphi_n^{(1)}(E_A^{(2)}+U_0^{(2)}V^{(2)}(U_0^{(2)})^{-1}-\lambda_2)\,\phi_N^{(2)}\|
\end{equation}
$$\leq \;\|\phi_N^{(2)}\|\;\epsilon\;+\;\|\varphi_n^{(1)}\|\;\epsilon$$
by assumption for $n$ sufficiently large.

We will now show that the remainder $r_1$ in (\ref{6.3}) is $\frac{1}{n}$-bounded.
This is equivalent  to proving  the following estimates,
$$|(\psi,U_0^{(1)}V^{(1)}(U_0^{(1)})^{-1}\;\psi_n)|\;\leq\; \frac{c(B_0)}{n}\;\|\psi\|\;\|\psi_n\|$$
\begin{equation}\label{6.4}
|(\psi,U_0^{(1)}U_0^{(2)} V^{(12)}(U_0^{(1)}U_0^{(2)})^{-1}\, \psi_n)|\;\leq\; \frac{c(B_0)}{n}\;\|\psi\|\;\|\psi_n\|
\end{equation}
for all  $\psi \in (C_0^\infty({\Bbb R}^3) \otimes {\Bbb C}^4)^2,$
with $c(B_0)$ some constant.
Using that $\varphi_n^{(1)}$ is localized outside the ball  $B_n(0)$ we define a smooth auxiliary function $\chi_1\in C^\infty({\Bbb R}^3)$ mapping to $[0,1]$,
\begin{equation}\label{5.10aa}
\chi_1\left(\frac{\bfx_1}{n}\right)\;=\;\left\{ \begin{array}{cc}
0,& x_1<Cn/2\\
1,& x_1\geq Cn,
\end{array}\right.
\end{equation}
and set $C=1.$
Then $\varphi_n^{(1)} = \chi_1 \varphi_n^{(1)} $ and we  have
$$|(\psi,U_0^{(1)} V^{(1)}(U_0^{(1)})^{-1}\,\chi_1\varphi_n^{(1)}\phi_N^{(2)})|\;\leq\; |((U_0^{(1)})^{-1} \psi,\,(\frac{\gamma}{x_1}\,\chi_1)\;(U_0^{(1)})^{-1}\varphi_n^{(1)} \phi_N^{(2)})|$$
\begin{equation}\label{6.5}
+\;|((U_0^{(1)})^{-1}\psi,\,\frac{\gamma}{x_1}\;[(U_0^{(1)})^{-1},\chi_1]\;\varphi_n^{(1)}\phi_N^{(2)})|.
\end{equation}
The first term is bounded by $2c/n$ since $\supp \chi_1$ requires $x_1\geq n/2.$ 
According to the note below (\ref{4.3i}) the second term can be estimated by 
$\frac{c}{n}\,\|\psi\|\,\|\varphi_n^{(1)}\|\,\|\phi_N^{(2)}\|.\;$

For the two-particle potential we define $\chi_{1}(\frac{\bfx_1-\bfx_2}{n})$ as in (\ref{5.10aa}) with $C=\frac12.$
Since $\phi_N^{(2)} \in C_0^\infty({\Bbb R}^3) \otimes {\Bbb C}^4$ there exists an $R_0>0$ such that $x_2<R_0$ on $\supp \phi_N^{(2)}.$
If one chooses $n$ such that  $n>2R_0,$ then $|\bfx_1-\bfx_2|\geq x_1-x_2>\frac{n}{2}$ on $\supp \psi_n.$ Thus $\varphi_n^{(1)} \phi_N^{(2)} =  \chi_{1} \,\varphi_n^{(1)} \phi_N^{(2)} $. In the decomposition
$$|(\psi,U_0^{(1)}U_0^{(2)}V^{(12)}(U_0^{(1)}U_0^{(2)})^{-1}\;\chi_{1}\,\varphi_n^{(1)} \phi_N^{(2)})|\;\leq\; |((U_0^{(1)}U_0^{(2)})^{-1}\psi,\frac{e^2}{|\bfx_1-\bfx_2|}\;\chi_{1}$$
$$\cdot(U_0^{(1)}U_0^{(2)})^{-1}\varphi_n^{(1)} \phi_N^{(2)})|\;
+\;|((U_0^{(1)}U_0^{(2)})^{-1}\psi,\frac{e^2}{|\bfx_1-\bfx_2|}\,\left\{ \,[(U_0^{(1)})^{-1},\chi_{1}]\,(U_0^{(2)})^{-1}\right.$$
\begin{equation}\label{6.6}
\left.+\;(U_0^{(1)})^{-1}\,[(U_0^{(2)})^{-1},\chi_{1}]\,\right\}\;\varphi_n^{(1)} \phi_N^{(2)})|
\end{equation}
the operator in the first term is $1/n$-bounded since $\chi_{1} \neq 0$ only if $\frac{1}{|\bfx_1-\bfx_2|}\,\leq 4/n.\;$
The operator containing $\frac{1}{|\bfx_1-\bfx_2|}\,[(U_0^{(1)})^{-1},\chi_{1}]$ 
is $1/n$-bounded according to  the note below (\ref{4.3i}).
For the last term we use a decomposition  and subsequent estimate as indicated in (\ref{B.4i}) and (\ref{B.5i}).
This proves the assertion (\ref{6.4}) and therefore,
\begin{equation}\label{6.8}
\|r_1\,\psi_n\|\;\leq\; \frac{c(B_0)}{n}\;\|\psi_n\|\;<\;\epsilon
\end{equation}
for $n$ sufficiently large. Hence the sequence $(\psi_n)_\nin$ obeys $\|(h_2^{BR}-\lambda)\,\psi_n\|<2\epsilon.$

The consideration of the antisymmetry of the sequence 
as well as its normalizability for sufficiently
large $n$ can be done in the same way as in the absence of a magnetic field \cite{Jaku2}.
Collecting results, this shows that  a subsequence of $(\cA \psi_n)_\nin$ is a Weyl sequence for $\lambda$
and verifies that $\lambda \in \sigma_{ess}(h_2^{BR}).$

\section{Proof of Theorem 1 (hard part)}
\setcounter{equation}{0}

Let us introduce the Ruelle-Simon partition of unity
$(\phi_j)_{j=0,1,2}\in C^\infty({\Bbb R}^6)$ subordinate to the two-cluster decompositions (\ref{5.2}) \cite{CFKS,LSV}. It is defined on the unit sphere and has the following properties,
$$\sum_{j=0}^2 \phi_j^2\;=\;1,\qquad \phi_j(\lambda \bfx)\;=\;\phi_j(\bfx)\quad \mbox{ for } x=1\;\mbox{ and } \lambda \geq 1,$$
$$\supp\phi_j\,\cap\,{\Bbb R}^6 \backslash B_1(0)\,\subset \{ \bfx \in {\Bbb R}^6 \backslash B_1(0):\,
|\bfx_1-\bfx_2|\geq Cx\mbox{ and } x_j\geq Cx\},\;\; j=1,2,$$
\begin{equation}\label{7.5}
\supp\phi_0\,\cap\,{\Bbb R}^6 \backslash B_1(0)\,\subset \,\{\bfx \in {\Bbb R}^6 \backslash B_1(0):\,x_k\geq Cx\;\;\forall\;k\in \{1,2\}\,\},
\end{equation}
where $\bfx=(\bfx_1,\bfx_2),\;\;x=|\bfx|\;$ and $C$ is a positive constant.
The (IMS-type) localization formula for  $h_2^{BR}$ is written in the following way 
\begin{equation}\label{7.6}
(u,h_2^{BR}\;u)\;=\;\sum_{j=0}^2 (\phi_j u, h_2^{BR}\,\phi_j u)\,-\,\sum_{j=0}^2 (\phi_j u, [h_2^{BR},\phi_j]\;u).
\end{equation}
For reasons which will become clear shortly we can  assume that the support of $u$ is outside the ball $B_R(0).$

First we show that in this case the commutator in the last term of (\ref{7.6}) tends to zero as $R \rightarrow \infty$. Explicitly, we have to prove
$$|\sum_{j=0}^2 (\phi_j u,[E_A^{(k)},\phi_j]\;u)|\;\leq\; \frac{c(B_0)}{R^2}\;\|u\|^2$$
\begin{equation}\label{7.7}
|(\phi_j u,[U_0^{(k)}V^{(k)}(U_0^{(k)})^{-1},\phi_j]\;u)|\;\leq\;\frac{c(B_0)}{R}\;\|u\|^2
\end{equation}
$$|(\phi_j u,[U_0^{(1)}U_0^{(2)}\,V^{(12)}\,(U_0^{(1)}U_0^{(2)})^{-1},\phi_j]\;u)|\;\leq\; \frac{c(B_0)}{R}\;\|u\|^2,$$
where $c(B_0)$ is a generic constant depending on the magnetic field.
Let us introduce the smooth  function $\chi \in C^\infty({\Bbb R}^6)$ mapping to $[0,1]$ by means of
\begin{equation}\label{7.8}
\chi\left( \frac{\bfx_1}{R},\frac{\bfx_2}{R}\right)\;=\;\left\{ \begin{array}{cc}
0,& x<R/2\\
1,&x\geq R
\end{array}\right.
\end{equation}
where $\bfx=(\bfx_1,\bfx_2),\;\;x=\sqrt{x_1^2+x_2^2}$ and $R>2.$ Then $\chi$ is unity on the support of $u$.
The resulting property $u = \chi \,u$ allows us to replace in (\ref{7.6}) the commutator  
form $(\phi_j u,[\cO,\phi_j]\,u)$ with $(\phi_j u,[\cO,\phi_j\chi]\,u)$ where the (arbitrary) operator $\cO$ can be identified with a constituent of $h_2^{BR}$.
For the kinetic energy operator of particle 1 (or for any single-particle operator $\cO$) one has the identity \cite{FLS}
$$\sum_{j=0}^2 (\phi_j u,[E_A^{(1)},\phi_j\chi]\,u)$$
\begin{equation}\label{7.9}
=\;-\frac{1}{2}\;\sum_{j=0}^2 \int_{{\Bbb R}^6} d\bfx \int_{{\Bbb R}^3}d\bfx'_1\;\overline{u}(\bfx)\,E_A^{(1)}(\bfx_1,\bfx'_1)\,\left[ (\phi_j \chi)(\bfx'_1,\bfx_2)-(\phi_j\chi)(\bfx)\right]^2\,u(\bfx'_1,\bfx_2).
\end{equation}

From the mean value theorem  we obtain
\begin{equation}\label{7.10}
\left| (\phi_j\chi)(\bfx'_1,\bfx_2)-(\phi_j \chi)(\bfx_1,\bfx_2)\right|^2\;\leq\; 
 \left| (\bfx_1-\bfx'_1)\;(\bfnabla_{\bfx_1} \phi_j\chi)(\bfxi,\bfx_2)\,\right|^2
\end{equation}
with $\bfxi$ on the line between $\bfx_1$ and $\bfx'_1$.
Since $\chi$ is supported outside $B_{R/2}(0)$ with $R/2>1$ the function $\phi_j$ obeys the scaling property
$\phi_j(\bfx)=\phi_j(\frac{\bfx}{R/2})$ from  the first line of  (\ref{7.5}) on $\supp \chi$.
Furthermore, $\phi_j$ and $\chi$ have a bounded derivative
since $\chi'\in C_0^\infty({\Bbb R}^6)$ and since $\phi_j \in C^\infty({\Bbb R}^6)$ is defined on the compact unit sphere, being homogeneous of
degree zero outside the unit ball.
Therefore we have $|\bfnabla_{\bfx_1}\phi_j\chi)(\bfxi,\bfx_2)|\,\leq c/R.$

With Lemma \ref{l4} for $E_A^{(1)}(\bfx_1,\bfx'_1),$ the $1/R^2$-boundedness of the kernel of the operator in (\ref{7.9}) is established by the Schur test (\ref{3.16}) (using $f=1$),
\begin{equation}\label{7.11}
I(\bfx_1,\bfx_2)\;=\;\frac12 \sum_{j=0}^2 \int_{{\Bbb R}^3} d\bfx'_1\;\frac{C(B_0)}{|\bfx_1-\bfx'_1|^4}\;e^{-(m-\epsilon)|\bfx_1-\bfx'_1|}\;|\bfx_1-\bfx'_1|^2\;\frac{c^2}{R^2}\;
\leq\; \frac{\tilde{C}(B_0)}{R^2}.
\end{equation}
This proves the first inequality of (\ref{7.7}).
\vspace{0.2cm}

For the single-particle potential contribution to $[h_2^{BR},\phi_j\chi]$ we have
$$[U_0^{(k)}V^{(k)}(U_0^{(k)})^{-1},\phi_j\chi]\;u\;=\;-\gamma\;[U_0^{(k)},\phi_j\chi]\,\frac{1}{x_k}\;(U_0^{(k)})^{-1}\;u$$
\begin{equation}\label{7.12}
-\gamma\,U_0^{(k)}\,\frac{1}{x_k}\;[(U_0^{(k)})^{-1},\phi_j\chi]\;u.
\end{equation}
The $\frac{1}{R}$-boundedness of these two terms is guaranteed by (\ref{4.3i}) and the note below.

The two-particle commutator can be treated in the same way, by
using
(\ref{B.4i}) and (\ref{B.5i}) together with the note following (\ref{4.3i}). This establishes the remaining inequalities of (\ref{7.7}).

In a next step we employ Persson's theorem (see e.g. \cite[Thm 3.12]{CFKS})
stating that
\begin{equation}\label{7.17}
\inf \sigma_{ess}(h_2^{BR})\;=\;\lim_{R \rightarrow \infty}\;\inf_{\|u\|=1}\, (u,h_2^{BR}\;u)
\end{equation}
if $u \in \cA(C_0^\infty({\Bbb R}^6 \backslash B_R(0))\otimes ({\Bbb C}^4)^2).\;$
The assumptions for Persson's theorem to hold are the relative form boundedness of the potential with respect to the
kinetic energy operator (proven in Lemma \ref{l1})
and the existence of a Weyl sequence $(u_n)_\nin$ to $\lambda \in \sigma_{ess}(h_2^{BR})$ where $u_n$ is supported outside the ball $B_n(0)$.
The proof of the latter item is given in Appendix B.

Inserting (\ref{7.6}) with (\ref{7.7}) into (\ref{7.17}) we obtain
\begin{equation}\label{7.18}
\inf\sigma_{ess}(h_2^{BR})\;=\;\lim_{R \rightarrow \infty}\;\inf_{\|u\|=1}\left\{ \sum_{j=0}^2 (\phi_ju,\,(E_{A,tot}+a_j)\;\phi_ju)\;
+\;\sum_{j=0}^2 (\phi_ju,r_j \,\phi_ju)\right\}
\end{equation}
where $h_2^{BR}\,=E_{A,tot}+a_j+r_j\;$ was used.

In the final step it remains to show that the second term in the curly brackets of  (\ref{7.18}) also tends uniformly to zero as $R \rightarrow \infty$.
If this is true then, recalling the definition (\ref{5.3a}) of $\Sigma_0$, one can estimate
$$\inf \sigma_{ess}(h_2^{BR})\;=\;\lim_{R \rightarrow \infty}\;\inf_{\|u\|=1}\sum_{j=0}^2 (\phi_ju,\,(E_{A,tot}+a_j)\,\phi_ju)$$
\begin{equation}\label{7.19}
\geq\; \lim_{R \rightarrow \infty} \;\inf_{\|u\|=1} \sum_{j=0}^2 \Sigma_0\;(\phi_ju,\phi_ju)\;=\;\Sigma_0
\end{equation}
which completes the proof of the hard part.

To provide the missing link let us start by taking $j=1$ and consider the single-particle contribution to $r_1.$

If in the auxiliary function $\chi_1(\bfx_1/R)$ from (\ref{5.10aa})
one takes  $C$ equal to the constant from the partition of unity (\ref{7.5}) and $R>1$ then $\chi_1=1$  on the support of $\phi_1u$ (note that $\supp \phi_1$ and $\supp u$ require $x_1\geq Cx$ and $x\geq R$, respectively).

We decompose
$$|(\phi_1u,U_0^{(1)}V^{(1)}(U_0^{(1)})^{-1}\,\chi_1\phi_1u)|\;\leq\;|((U_0^{(1)})^{-1}\phi_1u,\,\frac{\gamma}{x_1}\,\chi_1(U_0^{(1)})^{-1}\,\phi_1u)|$$
\begin{equation}\label{7.20}
+\;|((U_0^{(1)})^{-1}\,\phi_1u,\,\frac{\gamma}{x_1}\;[(U_0^{(1)})^{-1},\chi_1]\;\phi_1u)|.
\end{equation}
The first contribution can be estimated by $\frac{2\gamma}{CR}\,\|\phi_1u\|^2\;$
since $\supp \chi_1$ requires $x_1\geq CR/2.\;$
The $\frac{1}{R}$-boundedness of the second contribution follows from the note below (\ref{4.3i}).

For handling the two-particle contribution to $r_1$ we again introduce the function $\chi_1((\bfx_1-\bfx_2)/R)$ from (\ref{5.10aa}), its
argument being now the difference between the single-particle coordinates.
Again, $\chi_1=1$ on the support of $\phi_1u$. With a (\ref{7.20})-type
decomposition (where $U_0^{(1)}$ is replaced by $U_0^{(1)}U_0^{(2)}$ and $\gamma/x_1$ by $e^2/|\bfx_1-\bfx_2|)$
it is easy to see that the first term is bounded by $2e^2/(CR)\,\|\phi_1u\|^2.\;$
The second contribution is given by
$$\left| ((U_0^{(1)}U_0^{(2)})^{-1}\phi_1u,\frac{e^2}{|\bfx_1-\bfx_2|}\left\{[(U_0^{(1)})^{-1},\chi_1](U_0^{(2)})^{-1}\right.\right.$$
\begin{equation}\label{7.21}
\left.\left.+(U_0^{(1)})^{-1}[(U_0^{(2)})^{-1},\chi_1]\,\right\}\phi_1u)\right|.
\end{equation} 
The $\frac{1}{R}$-boundedness of this contribution is established by the note following (\ref{4.3i}) (with the help
of (\ref{B.4i})- and (\ref{B.5i})-type decompositions).

The case $j=2$ follows from the symmetry of $h_2^{BR}$ under particle exchange.

For $j=0$ we have $r_0\,=\sum\limits_{k=1}^2 U_0^{(k)}V^{(k)}(U_0^{(k)})^{-1}\;$ and we introduce also here the function $\chi_1(\bfx_k/R)$ from (\ref{5.10aa}).
Since $\supp \phi_0u\subset \{x\in {\Bbb R}^6:\,x\geq R,\;x_1\geq CR,\,x_2\geq CR\},\;$ we have $\phi_0u=\chi_1\,\phi_0u$ for both values of $k$.
The proof of the $1/R$-boundedness of $(\phi_0u,r_0\,\phi_0u)$ is therefore the same as for the $j=1$ single-particle case.
In conclusion, this shows that  (\ref{7.18}) 
reduces to (\ref{7.19}) which completes the proof.

We remark that the present proof is valid for field strengths $0\leq B_0<\infty$ and thus covers the case $\bfA = {\bf 0}$ as well.

\section*{Appendix A\quad (Integral formulae)}
\renewcommand{\theequation}{\Alph{section}.\arabic{equation}}
\setcounter{equation}{0}
\setcounter{section}{1}

For convenience we cite some general formulae. We have \cite[p.340]{G}
\begin{equation}\label{A.6a}
\int_0^\infty dt\; t^\nu\;e^{-\gamma t}\;e^{-\beta/t}\;=\;2\;\left( \frac{\beta}{\gamma}\right)^{\frac{\nu+1}{2}}\;K_{\nu+1}(2\sqrt{\beta \gamma}\,),\qquad \beta,\,\gamma >0.
\end{equation}

Moreover \cite[p.705]{G},
$$\int_0^\infty dt\;\frac{t^{2\mu+1}}{(\sqrt{t^2+a^2})^\nu}\;K_\nu(\alpha \sqrt{t^2+a^2}\,)\;=\;\frac{2^\mu\Gamma(\mu+1)}{\alpha^{\mu+1}a^{\nu-\mu-1}}\;K_{\nu-\mu-1}(\alpha a),$$
\begin{equation}\label{A.6b}
\quad \alpha>0,\;\;a>0,\;\;\mu>-1.
\end{equation}

We also provide the asymptotic formulae for the modified Bessel functions \cite[p.374]{AS},
\begin{eqnarray}\label{A.8}
K_0(z)\;\sim\;-\ln z,& K_1(z) \;\sim\; \frac{1}{z},& K_\nu(z)\;\sim\; \frac{\Gamma(\nu) 2^{\nu-1}}{z^\nu},\;\;\nu>0,\quad \mbox{for } z \rightarrow 0\nonumber\\
K_\nu(z)\;\sim\;\sqrt{\frac{\pi}{2z}}\;e^{-z},&\nu\geq 0,& \mbox{for } z \rightarrow \infty 
\end{eqnarray}
and recall that $K_{-\nu}(z)\,=K_\nu(z)$, as well as $K_{\nu+1}(z)\,=\,K_{\nu-1}(z)+\;\frac{2\nu}{z}\;K_\nu(z).$

\section*{Appendix B\quad (Existence of Weyl sequence outside balls)}
\renewcommand{\theequation}{\Alph{section}.\arabic{equation}}
\setcounter{equation}{0}
\setcounter{section}{2}

\begin{lemma}\label{l6a}
Let $h_2^{BR}=E_{A,tot}+w$, $\bfB$ a constant magnetic field, and let $w$ be relatively form bounded with respect to $E_{A,tot}$.
If $\lambda \in \sigma_{ess}(h_2^{BR})$ there exists a Weyl sequence $({u}_n)_\nin$ to $\lambda$ with the additional property $u_n \in \cA(C_0^\infty({\Bbb R}^6\backslash B_n(0))\otimes ({\Bbb C}^4)^2),$ where $B_n(0)$ is a ball of radius $n$ centered at the origin.
\end{lemma}

The proof follows closely the one given in \cite{CFKS} and in \cite{Jaku2}. For $\lambda \in \sigma_{ess}(h_2^{BR})$ there exists a Weyl sequence $\psi_n 
\in \cA(C_0^\infty({\Bbb R}^3) \otimes {\Bbb C}^4)^2,\;\;\|\psi_n\|=1\;$ with $\psi_n \stackrel{w}{\rightharpoonup} 0$ and $\|(h_2^{BR}-\lambda)\,\psi_n\|\,\rightarrow 0$ as $\ninfi.$
Thus for any $\epsilon := \frac{1}{n}>0$ there exists some $N(n)>n$ such that
$\|(h_2^{BR}-\lambda)\psi_{N(n)}\|\,<\epsilon.$

In order to construct the Weyl sequence $({u}_n)_\nin$, which consists of functions
 localized outside $B_n(0)$, we define a smooth function $\chi_{0n} \in C_0^\infty({\Bbb R}^6)$ which is symmetric (with respect to the interchange of $\bfx_1$ and $\bfx_2$) and maps to $[0,1]$, 
\begin{equation}\label{B.1}
\chi_{0n}(\bfx):=\;\chi_0\left(\frac{\bfx}{n}\right)\;=\;\left\{ \begin{array}{cc}
1,& x\leq n\\
0,& x>2n
\end{array} \right. .
\end{equation}
Then we claim that a subsequence of $(\chi_n\psi_{N(n)})_\nin$ with $\chi_n :=1-\chi_{0n}$
is the desired sequence $(u_n)_\nin$. 

In order to show that $\|(h_2^{BR}-\lambda)\, \chi_n\psi_{N(n)}\|\,\rightarrow 0$ as $\ninfi$, we   decompose
\begin{equation}\label{B.2}
\|(h_2^{BR}-\lambda)\,\chi_n\psi_{N(n)}\|\;\leq\; \|\chi_n\,(h_2^{BR}-\lambda)\,\psi_{N(n)}\|\;+\;\|\,[h_2^{BR},\chi_{0n}]\,\psi_{N(n)}\|.
\end{equation}
The first term goes to zero for $\ninfi$ by assumption since $\chi_n$ is bounded.
As regards the second term, the kinetic energy contribution $\|\,[E_A^{(k)},\chi_{0n}]\,\psi_{N(n)}\|\,\leq \frac{c(B_0)}{n}\;$ by Lemma \ref{l5} since $\chi_{0n}$ as a $C_0^\infty$-function has bounded derivatives.
The coordinate of the second particle, $\bfx_{\bar{k}}$ (with $\bar{k} \in \{1,2\}\backslash k),$ in $\chi_{0n}$ can be treated as a parameter.

For the contribution from the single-particle potential we have
\begin{equation}\label{B.3i}
[U_0^{(k)}V^{(k)}(U_0^{(k)})^{-1},\chi_{0n}]\;=\;-\gamma\,[U_0^{(k)},\chi_{0n}]\,\frac{1}{x_k}\,(U_0^{(k)})^{-1}\,
-\,\gamma U_0^{(k)}\,\frac{1}{x_k}\,[(U_0^{(k)})^{-1},\chi_{0n}].
\end{equation}
Each of the two terms is $\frac{1}{n}$-bounded according to (\ref{4.3i}) and the note below.

The commutator with the two-particle potential is written in the following way,
$$[U_0^{(1)}U_0^{(2)}V^{(12)}(U_0^{(1)}U_0^{(2)})^{-1},\chi_{0n}]$$
$$=\;e^2\,[U_0^{(1)},\chi_{0n}]\,\frac{1}{|\bfx_1-\bfx_2|}\left( |\bfx_1-\bfx_2|\,U_0^{(2)}\,\frac{1}{|\bfx_1-\bfx_2|}\right)(U_0^{(1)}U_0^{(2)})^{-1}
+\;e^2U_0^{(1)}$$
\begin{equation}\label{B.4i}
\cdot[U_0^{(2)},\chi_{0n}]\,\frac{1}{|\bfx_1-\bfx_2|}\,(U_0^{(1)}U_0^{(2)})^{-1}\,+\,e^2U_0^{(1)}U_0^{(2)}\frac{1}{|\bfx_1-\bfx_2|}\,[(U_0^{(1)})^{-1},\chi_{0n}]\,(U_0^{(2)})^{-1}
\end{equation}
$$+\;e^2U_0^{(1)}U_0^{(2)}\left(\frac{1}{|\bfx_1-\bfx_2|}\,(U_0^{(1)})^{-1}|\bfx_1-\bfx_2|\right)\,\frac{1}{|\bfx_1-\bfx_2|}\,[(U_0^{(2)})^{-1},\chi_{0n}].$$
All commutators (including the factor
 $\frac{1}{|\bfx_1-\bfx_2|}$) are $\frac{1}{n}$-bounded by Corollary \ref{c1i} to Lemma \ref{l7a} and by the note following (\ref{4.3i}).
The boundedness of the factors in round brackets follows from the decomposition
\begin{equation}\label{B.5i}
\frac{1}{|\bfx_1-\bfx_2|}\,(U_0^{(1)})^{-1}\,|\bfx_1-\bfx_2|\;=\;(U_0^{(1)})^{-1}\;+\;\frac{1}{|\bfx_1-\bfx_2|}\;[(U_0^{(1)})^{-1},\,|\bfx_1-\bfx_2|\,]
\end{equation}
where $|\bfx_1-\bfx_2|$ satisfies the requirements of Corollary \ref{c1i} for the function $g$.

It remains to prove that $\chi_{n} \psi_{N(n)}$ is normalizable for sufficiently large $n$.
To this aim we show that $\|\chi_{0n}\psi_{N(n)}\|\rightarrow 0$ for $\ninfi$.
Then $\;\|\chi_{n}\psi_{N(n)}\|\\=\,\|\psi_{N(n)}-\chi_{0n}\psi_{N(n)}\|\,\rightarrow \,1$ since $ \|\psi_{N(n)}\|\,=1$ for all $n$.
Due to  the semiboundedness of $h_2^{BR}$ there exists  $\mu>0$ such that $h_2^{BR}+\mu$ has a bounded inverse.
We estimate
$$\|\chi_{0n} \psi_{N(n)}\|\;=\;\|\chi_{0n}\,(h_2^{BR}+\mu)^{-1}\,[\,(h_2^{BR}-\lambda)\,+\,(\mu+\lambda)\,]\,\psi_{N(n)}\|$$
\begin{equation}\label{B.8}
\leq \;\|\chi_{0n}\,(h_2^{BR}+\mu)^{-1}\|\;\|\,(h_2^{BR}-\lambda)\,\psi_{N(n)}\|\;+\;|\mu+\lambda|\;\|\chi_{0n}\,(h_2^{BR}+\mu)^{-1}\,\psi_{N(n)}\|.
\end{equation}
The first summand is bounded by $\epsilon$ times a constant by assumption.
The second term tends to zero 
provided we can show that $\chi_{0n}(h_2^{BR}+\mu)^{-1}$ is a compact operator which turns the weakly convergent sequence $(\psi_{N(n)})_\nin$ into a strongly convergent one.
Consider the decomposition
$$\chi_{0n}\,(h_2^{BR}+\mu)^{-1}\,\psi_{N(n)}\;=\;\left\{\chi_{0n}\left( \sum_{k=1}^2 S_A^{(k)2}\right)^{-\frac{1}{4}}\right\}$$
\begin{equation}\label{B.9}
\cdot \left[\!\! \left( \sum_{k=1}^2 S_A^{(k)2}\!\right)^{\!\!\frac{1}{4}}\!\!\!\!(E_{A,tot}+\mu)^{-\frac12}\!\right]\!\!\left((E_{A,tot}+\mu)^\frac12 (h_2^{BR}+\mu)^{-\frac12}\right)(h_2^{BR}+\mu)^{-\frac12}\psi_{N(n)}.
\end{equation}
From the diamagnetic inequality-based relation (\ref{2.2c}) it immediately follows
that (for fixed $n$) the operator in curly brackets is compact since
$\chi_{0n} (\sum_{k=1}^2(p_k^2+m^2))^{-1/4}$ is compact.

For the boundedness of the operator in square brackets we invoke an estimate proven by Balinsky, Evans and Lewis \cite{BEL} for the Pauli operator (i.e. for the case $m=0$
in the lemma below).

\begin{lemma}\label{l6}
For a single particle let $S_A= \left[ (\bfp-e\bfA)^2+m^2\right]^\frac12\;$ and $E_A$ from (\ref{1.4a}) with $\bfA \in L_{2,loc}({\Bbb R}^3).$ Then the following estimate holds,
\begin{equation}\label{B.10}
E_A^2\;\geq\; \delta_m^2(B)\;S_A^2,
\end{equation}
where
\begin{equation}\label{B.10a}
\delta_m(B)\;=\;\inf_{\|f\|=1}\;\|\,(1-S_m^\ast S_m)\,f\|\;>\,0
\qquad \mbox{with } S_m:=\;(eB)^\frac12 (E_A^2+eB)^{-\frac12}
\end{equation}
and $f \in L_2({\Bbb R}^3) \otimes {\Bbb C}^4.$
\end{lemma}

We note that
\begin{equation}\label{B.9ii}
S_mS_m^\ast\;=\;eB^\frac12\;(E_A^2+eB)^{-1}\;B^\frac12\;\leq\; eB^\frac12 \;(m^2+eB)^{-1}\;B^\frac12\;<\;1
\end{equation}
(and hence also $S_m^\ast S_m <1)\;$ since for $m\neq 0$ zero modes are absent irrespective of $\bfB.$
Therefore for all $f$ with $\|f\|=1$ one has 
$\,1>\,(f,S_mS_m^\ast f)\,=\,\|S_m^\ast f\|^2$ and hence $1>\,\sup_f \|S_m^\ast f\|\,=\|S_m^\ast\|.\;$
Thus the proof of the lemma
can be copied from \cite{BEL}.
For a constant magnetic field $\bfB_0,\;\;S_m^\ast S_m\,\leq \frac{eB_0}{m^2+eB_0}.$ 
This leads to $\delta_m(B_0)\,\geq \,\frac{m^2}{m^2+eB_0}.\;$

The required boundedness of the operator in (\ref{B.9}) is based on  the existence of  a constant $c$ such that for $\varphi \in \cA(L_2({\Bbb R}^3) \otimes {\Bbb C}^4)^2,$
\begin{equation}\label{B.11}
\left\| \left( \sum_{k=1}^2 S_A^{(k)2}\right)^{\!\!\frac{1}{4}}\!\!\left( E_{A,tot}+\mu\right)^{-\frac12}\varphi\,\right\|^2\leq\;
\left\|\left( \sum_{k=1}^2 S_A^{(k)}\right)^{\!\!\frac12}\!\!(E_{A,tot} +\mu)^{-\frac12} \varphi\,\right\|^2
\stackrel{!}{\leq} \, c^2\,\|\varphi\|^2.
\end{equation}
Defining $\tilde{\varphi}:= (E_{A,tot}+\mu)^{-\frac12}\varphi$ this is equivalent to proving (for $\mu\geq 0)$
\begin{equation}\label{B.12}
 (\tilde{\varphi}, \sum_{k=1}^2 S_A^{(k)}\;\tilde{\varphi})\;\leq\; c^2\;(\tilde{\varphi},\left( E_{A,tot}+\mu\right) \;\tilde{\varphi}),
\end{equation}
which is assured by Lemma \ref{l6} with $c^2=\frac{1}{\delta_m(B)}.$

The boundedness of the next term in (\ref{B.9}) makes use of the relative form boundedness of the potential (\ref{2.5}).
Defining $\phi_n:= (h_2^{BR}+\mu)^{-\frac12}\tilde{\psi}_n$ with 
$\tilde{\psi}_n:= (h_2^{BR}+\mu)^{-\frac12}\,\psi_{N(n)},$ we have to show
\begin{equation}\label{B.14}
\|\,(E_{A,tot}+\mu)^\frac12(h_2^{BR}+\mu)^{-\frac12}\;\tilde{\psi}_n\|^2\;=\;(\phi_n,(E_{A,tot}+\mu)\;\phi_n)\;\leq\; c^2 \;(\phi_n,(h_2^{BR}+\mu)\,\phi_n).
\end{equation}
Using (\ref{2.5}) the r.h.s. of (\ref{B.14}) can be estimated,
$$(\phi_n, (h_2^{BR}+\mu)\,\phi_n)\;\geq\; (\phi_n,(E_{A,tot}+\mu)\,\phi_n)\,-\;|(\phi_n,w\,\phi_n)|$$
\begin{equation}\label{B.15}
\geq\; (1-c_0)\,(\phi_n, (E_{A,tot}+\mu)\,\phi_n)\;-\;C_1(B_0)\;\|\phi_n\|^2\;+\;c_0\mu\,(\phi_n,\phi_n)
\end{equation}
with $c_0:= \gamma\pi/2\,+e^2\pi/4\,<1.$
If $\mu$ is chosen larger than $C_1(B_0)/c_0$ then the terms proportional to $\|\phi_n\|^2$ can be dropped, such that the second line of (\ref{B.15}) is $\geq \,\frac{1}{c^2}\,(\phi_n, (E_{A,tot}+\mu)\,\phi_n)\;$ with
$c^2:= \frac{1}{1-c_0}.\;$
This proves (\ref{B.14}). 

To complete the proof of the compactness of $\chi_{0n}(h_2^{BR}+\mu)^{-1}$ we keep $n$ fixed.
From the discussion above there exists  to the given $\epsilon =1/n$ an $N(n)>n$ such that
$\|\chi_{0n}(\sum\limits_{k=1}^2 S_A^{(k)2})^{-1/4} \cdot B\,\psi_{N(n)}\|\,<\epsilon. \;\;B$ comprises the bounded operators in (\ref{B.9}) to the right of the one in curly brackets,
and $N(n)$ has to be chosen large enough to satisfy the previous
condition $\|(h_2^{BR}-\lambda)\psi_{N(n)}\|\,<\epsilon\,$ as well.
This proves $\|\chi_{0n}\psi_{N(n)}\|\,\leq\, c\,\epsilon +|\mu+\lambda|\,\epsilon\,$
and hence the normalizability of the Weyl sequence under consideration.
 
\section*{Appendix C\quad (Estimate for the kernel of $A_E$)}
\renewcommand{\theequation}{\Alph{section}.\arabic{equation}}
\setcounter{equation}{0}
\setcounter{section}{3}

For $\sqrt{2} A_E$ we use the representation
\begin{equation}\label{C.1}
\sqrt{\frac{E_A+m}{E_A}}\;=\;\lim_{t \rightarrow 0}\;\frac{1-e^{-t(E_A+m)}}{t\,\sqrt{E_A}\,\sqrt{E_A+m}}.
\end{equation}
With the integral formula (\ref{a.7}) for each of the factors
in the denominator we obtain, using the Taylor formula to first order,
$$\sqrt{\frac{E_A+m}{E_A}}\;=\;\lim_{t \rightarrow 0}\;\frac{1}{t\pi}\int_0^\infty \frac{dt'}{\sqrt{t'}}\int_0^\infty \frac{d\tau}{\sqrt{\tau}}\;e^{-\tau m}
\,\left\{ e^{-(\tau +t')E_A}\,
-\,e^{-tm}\,e^{-(\tau+t'+t)E_A}\right\}$$
\begin{equation}\label{C.2}
=\;-\frac{1}{\pi}\int_0^\infty \frac{dt'}{\sqrt{t'}}\int_0^\infty \frac{d\tau}{\sqrt{\tau}}\; e^{-\tau m}\left\{\frac{d}{d\tau}\left( e^{-(\tau +t')E_A}\right)\,-\,m\,e^{-(\tau + t')E_A}\right\}.
\end{equation}
As a next step we  estimate the kernel of $e^{-tE_A}$ and its derivative.
Using the integral representation (\ref{3.14b}) and estimating the  kernel
of $e^{-tE_A^2}$  with the help of  (\ref{a.6a}) leads to
\begin{equation}\label{C.3}
|e^{-tE_A}(\bfx,\bfx')|\;
\leq\;\frac{1}{\pi^2t^3}\int_0^\infty \tau\;d\tau\;e^{-\tau}\;e^{-t^2m^2/4\tau}\;e^{-(\bfx-\bfx')^2\tau/t^2}\left( 1\,+\,eB_0\;\frac{t^2}{2\tau}\right).
\end{equation}
From (\ref{A.6a}) one obtains, abbreviating $\xi:= \sqrt{t^2+(\bfx-\bfx')^2},$
\begin{equation}\label{C.4}
|e^{-tE_A}(\bfx,\bfx')|\;\leq\; \frac{m}{2\pi^2}\left\{ \frac{mt}{\xi^2}\;K_2(m\xi)\,+\,eB_0\;\frac{t}{\xi}\;K_1(m\xi)\right\}.
\end{equation}
For the derivative we have with $\varrho:= t^2/(4\tau),$
\begin{equation}\label{C.5}
\frac{d}{dt}\;e^{-tE_A}(\bfx,\bfx')\;=\;\frac{1}{\sqrt{\pi}}\int_0^\infty \frac{d\tau}{\sqrt{\tau}}\;e^{-\tau}\;\frac{d}{d\varrho} \left( e^{-\varrho E_A^2}(\bfx,\bfx')\right)\cdot \frac{t}{2\tau}.
\end{equation}
With  the help of (\ref{b.7}) and its estimate (\ref{b.8}) one obtains
$$\left| \frac{d}{dt}\;e^{-tE_A}(\bfx,\bfx')\right|\;\leq\; \frac{1}{2\pi^2t^2}\int_0^\infty d\tau\;e^{-\tau}\;e^{-m^2t^2/4\tau}\;e^{-(\bfx-\bfx')^2\tau/t^2}$$
\begin{equation}\label{C.6}
\cdot\,\left[ \frac{6\tau}{t^2}+5eB_0 +m^2+2eB_0(\bfx-\bfx')^2\frac{\tau}{t^2}+(\bfx-\bfx')^2\frac{4\tau^2}{t^4}+(eB_0m^2+2e^2B_0^2)\frac{t^2}{2\tau}\right]
\end{equation}
which, using the integral formula (\ref{A.6a}), reduces to
$$\left| \frac{d}{dt}\;e^{-tE_A}(\bfx,\bfx')\right|\;\leq\;\frac{1}{\pi^2}\left\{ [3+eB_0(\bfx-\bfx')^2]\;\frac{m^2}{2\xi^2}\,K_2(m\xi)\,+\,(5eB_0\right.$$
\begin{equation}\label{C.7}
\left. +m^2)\;\frac{m}{2\xi}\;K_1(m\xi)\,+\,\frac{m^3}{2\xi^3}(\bfx-\bfx')^2\;K_3(m\xi)\,+\,(\frac{eB_0m^2}{2}\,+e^2B_0^2)\;K_0(m\xi)\right\}.
\end{equation}
Insertion of (\ref{C.4}) and (\ref{C.7}) into (\ref{C.2}) with $t:=\tau+t'$ results in
$$\left| \sqrt{\frac{E_A+m}{E_A}}(\bfx,\bfx')\right|\;\leq\; \frac{1}{\pi^3}\int_0^\infty \frac{d\tau}{\sqrt{\tau}}\;e^{-\tau m}\int_0^\infty \frac{dt'}{\sqrt{t'}}\left\{ (3+eB_0(\bfx-\bfx')^2\frac{}{}\right.$$
\begin{equation}\label{C.8}
\cdot \frac{m^2}{2\xi^2}\;K_2(m\xi)\,
+\,(5eB_0+m^2)\;\frac{m}{2\xi}\;K_1(m\xi)\,+\,\frac{m^3}{2\xi^3}\,(\bfx-\bfx')^2\;K_3(m\xi)
\end{equation}
$$\left. +\,(\frac{eB_0m^2}{2}\,+e^2B_0^2)\;K_0(m\xi)\,+\,\frac{m^2}{2}\,(\tau + t')\left[ \frac{m}{\xi^2}\;K_2(m\xi)\,+\,eB_0\;\frac{K_1(m\xi)}{\xi}\right] \right\}.$$

Now we follow the strategy given below (\ref{4.15s}), i.e. we estimate $e^{-\tau m}\leq 1$ and $\xi^{-\nu} K_\nu(m\xi) \,\leq y^{-\nu}K_\nu(my)$ with $y:= \sqrt{\tau^2+t^{'2}+(\bfx-\bfx')^2}$ in (\ref{C.8}). Then
 the double integral can be performed analytically by  (\ref{A.6b}).
 The result is
$$\left| \sqrt{\frac{E_A+m}{E_A}}(\bfx,\bfx')\right|\;\leq\;
c_0\,\frac{K_{3/2}(m|\bfx-\bfx'|)}{|\bfx-\bfx'|^{3/2}}\,+\, c_1(B_0)\;|\bfx-\bfx'|^{1/2}\;K_{3/2}(m|\bfx-\bfx'|)$$
$$+\,c_2(B_0)\;\frac{K_{1/2}(m|\bfx-\bfx'|)}{|\bfx-\bfx'|^{1/2}}\;
+\;c_3\;\frac{K_{5/2}(m|\bfx-\bfx'|)}{|\bfx-\bfx'|^{1/2}}$$
\begin{equation}\label{C.10}
+c_4(B_0)|\bfx-\bfx'|^{1/2}K_{1/2}(m|\bfx-\bfx'|)+2c_5\frac{K_1(m|\bfx-\bfx'|)}{|\bfx-\bfx'|}
+2c_6(B_0)K_0(m|\bfx-\bfx'|).
\end{equation}
From (\ref{C.10}) follows the estimate
\begin{equation}\label{C.11}
|A_E(\bfx,\bfx')|\;=\;\left| \sqrt{\frac{E_A+m}{2 E_A}}(\bfx,\bfx')\right|\;\leq\; \frac{\tilde{c}(B_0)}{|\bfx-\bfx'|^3}\;e^{-(m-\epsilon)|\bfx-\bfx'|}
\end{equation}
for $\epsilon \in (0,m).\;$

We note that by (\ref{C.4}) this proof provides an estimate for the heat kernel $e^{-tE_A}(\bfx,\bfx')$ of the Brown-Ravenhall operator for a free electron in a constant magnetic field.

\section*{Appendix D\quad (Erratum to Reference 14)}
\renewcommand{\theequation}{\thesection.\arabic{equation}}
\setcounter{equation}{0}
\setcounter{section}{3}

We collect the changes due to the replacement of $(\varphi,(\bfp-e\bfA)^2\,\varphi) \geq (\varphi, p^2\varphi)\;$ (which is not generally valid) by (\ref{2.2c}). We emphasize that this only affects the proofs, but not the results of \cite{Jaku1}.

Eqs. (3.1), (3.2) should be deleted. Instead, one has
$$\|\frac{1}{x}\;\varphi\|^2\;\leq\;4\; \|\sqrt{(\bfp-e\bfA)^2 +m^2}\;\varphi\|^2,$$
\begin{equation}
(\varphi, \frac{1}{x}\,\varphi)\;\leq\; \frac{\pi}{2}\;(\varphi, \sqrt{(\bfp-e\bfA)^2 +m^2}\;\varphi),
\end{equation}
valid for $m\geq 0.$

Eq. (3.3) should be replaced by the two separate estimates,
\setcounter{equation}{2}
$$\mbox{tr} \left[ p^2\,+\,\frac{e\bfsigma\bfB}{\mu}\right]_-^d\;\leq\; 2\,L_{d,3}\int_{{\Bbb R}^3}\left( \frac{e|\bfB|}{\mu}\right)^{d+\frac{3}{2}}d\bfx,$$
\begin{equation}
\mbox{tr }[\mu (\bfp-e\bfA)^2+e\bfsigma \bfB]_-^d\;\leq\; 2\,\mu^d\,L_{d,3}\int_{{\Bbb R}^3}\left( \frac{e|\bfB|}{\mu}\right)^{d+\frac{3}{2}}d\bfx.
\end{equation}

{\noindent The paragraph on p.7512 starting with 'The  strategy ...' 
should be replaced by:}

\setcounter{section}{5}
\setcounter{equation}{13}
The strategy to show the compactness of $K$ is to start with the operator $K_0:= \chi_0(p^2+m^2)^{-\frac12}$ which is compact as a product of bounded functions
$f(\bfx), \,g(\bfp),$ each of which tending to zero as $x$, respectively $p$, go to infinity (see, e.g., [31, Lemma 7.10]).
Then $K_1:= \chi_0 [(\bfp-e\bfA)^2+m^2]^{-\frac12}$ is also compact \cite[p.117]{CFKS}. In the following, bounded operators $\cO_1,\cO_2$ are
constructed such that $K_1 \cdot \cO_1 \cdot \cO_2=K.$

Let $\cO_1:= \sqrt{(\bfp-e\bfA)^2+m^2}/E_A.$
For showing the boundedness of $\cO_1$ let $\psi := E_A^{-1}\varphi.$
Then from (4.6),
$$\|\cO_1 \varphi\|^2\;=\; (\psi, ((\bfp-e\bfA)^2+m^2)\,\psi)\;\leq
\; (\psi,(E_A^2+e|\bfB|)\,\psi)$$
\begin{equation}
\leq\; \frac{1}{1-\kappa e}\;\|\varphi\|^2\,+\,\frac{eC_\kappa}{1-\kappa e}\,\|E_A^{-1}\|^2\;\|\varphi\|^2,
\end{equation}
the rhs being obviously bounded. With $\cO_2:= E_A \Lambda_{A,+}(H_0+\mu)^{-1}\leq 1\;$ (as shown above),
we have proven the compactness of $K$.

\section*{Acknowledgment}

I would like to thank L.Erd\"{o}s for clarifying discussions
and S.Morozov for critical comments.

\vspace{1cm}

\end{document}